\documentclass[fleqn,10pt]{wlscirep}
\usepackage[utf8]{inputenc}
\usepackage[T1]{fontenc}
\usepackage{gensymb}
\usepackage{graphicx}
\usepackage{subcaption}
\usepackage{multirow}
\usepackage{amsmath}
\usepackage{array}
\usepackage{hyperref}
\title{Universality  Class Transition Across the Helimagnetic Ordering in Te-doped Cu$_2$OSeO$_3$}

\author[1,2$\dagger$]{A.J. Ferguson}
\author[1,3,4$\dagger$]{M. V\'as}
\author[1,3]{E. J. Vella}
\author[5]{Md. F. Pervez}
\author[6]{E. P. Gilbert}
\author[5*]{C. Ulrich}
\author[1,3*]{S. Yick}
\author[1,3*]{T. S\"ohnel}

\affil[1]{School of Chemical Sciences, The University of Auckland, Auckland, 1142, New Zealand}
\affil[2]{Department of Physics, University of Fribourg, Fribourg, 1700, Switzerland}
\affil[3]{MacDiarmid Institute for Advanced Materials and Nanotechnology, Wellington, 6140, New Zealand}
\affil[4]{Australian Institute of Nuclear Science and Engineering, Lucas Heights, New South Wales, 2234, Australia}
\affil[5]{School of Physics, University of New South Wales, Sydney, New South Wales, 2052, Australia}
\affil[6]{Australian Centre for Neutron Scattering, ANSTO, Lucas Heights, New South Wales, 2234, Australia}

\affil[$\dagger$]{These authors contributed equally}
\affil[*]{corresponding authors samuel.yick@auckland.ac.nz, c.ulrich@unsw.edu.au, and t.soehnel.@auckland.ac.nz}

\begin{abstract}
Cu$_2$OSeO$_3$ is a multiferroic insulating chiral magnet which exhibits various magnetic orderings at different conditions. The positions of these magnetic phase transitions are known to be sensitive to chemical substitution. Here, we present a universality class analysis of the Cu$_2$OSe$_{1-x}$Te$_x$O$_3$ with $(0\leq x \leq 0.1)$. Tellurium is a non-magnetic ion which, upon substitution into the selenium positions of the structure, applies a positive chemical pressure and expands the crystal lattice. Our SANS and magnetometry data indicate that Te inclusion lowers the paramagnetic to helical ordering temperature and the critical field required for the conical to field polarised phase transition. By using the Heat-map Modified Iteration Method that evaluates critical behaviour on both sides of the transition, we show that the Heisenberg to Ising universality class transition of the initial ordering is robust to internal chemical pressure. Additionally, we attribute the decreases with doping in both critical temperature and critical field to be due to decreases in the strength of the Dzyaloshinskii-Moriya and Symmetric Exchange Interactions. 
\end{abstract}
\begin{document}
\setlength{\parindent}{0pt}

\flushbottom
\maketitle
%
%
\thispagestyle{empty}

\section*{Introduction}
Magnetic skyrmions are stable configurations of spins, characterised by their winding number which is a topological invariant. Since the discovery of magnetic skyrmions in MnSi more than a decade ago\cite{muhlbauer_skyrmion_2009}, B20 chiral magnets have been investigated both for their ability to host magnetic skyrmions, as well as for potential spintronics applications\cite{jonietz_spin_2010, huang_situ_2018, tokura_magnetic_2021}. This class of materials crystallise in the \textit{P$2_13$} space group, with the non-centrosymmetric nature of this structure allowing for a non-zero Dzyaloshinskii-Moriya Interaction (DMI)\cite{moriya_anisotropic_1960}. The resulting competition between symmetric exchange interactions (SEI), DMI, and magnetocrystalline anisotropies (MCA) leads to material systems with many shared chiral magnetic phases\cite{kanazawa_noncentrosymmetric_2017}. Commonly investigated materials in this family include the itinerant magnets MnSi\cite{causer_small-angle_2023, pfleiderer_skyrmion_2010}, Fe$_{1-x}$Co$_x$Si\cite{pfleiderer_skyrmion_2010, adams_skyrmion_2010, zhang_spin-dimensionality_2016}, FeGe\cite{yu_real-space_2010, yu_near_2011, zhang_critical_2016}, and the multiferroic insulator Cu$_2$OSeO$_3$\cite{seki_observation_2012,adams_long-wavelength_2012, flores_scaling_2023}. The electrically insulating and multiferroic properties of Cu$_2$OSeO$_3$ are unique amongst B20 skyrmion hosting materials. This makes it an ideal candidate for applications such as skyrmion based racetrack memory devices as it would avoid the joule heating issues of common electric current-based manipulation techniques\cite{kanazawa_noncentrosymmetric_2017}. Additionally, there are exchange anisotropy driven, low temperature skyrmions and tilted conical phases observed in Cu$_2$OSeO$_3$ that have not been observed in the other B20 skyrmion hosting materials\cite{halder_thermodynamic_2018, qian_new_2018}.\\ \\

Alongside the well known paramagnetic (PM), helimagnetic (HM), and conical phases that have been studied in B20 materials since the 1970's\cite{ludgren_helical_1970, ishikawa_helical_1976, bak_theory_1980}, B20 materials have also been shown to display a fluctuation-induced first order transition. The interactions between critical fluctuations have the effect of suppressing the expected continuous PM to HM transition, instead resulting in a continuous transition between the PM and a fluctuation disordered (FD) phase, followed by an induced first order transition from the FD phase to the HM phase. This is a phenomenon seen in many of the B20 skyrmion materials\cite{janoschek_fluctuation_2013, bauer_history_2016}, with different scenarios depending on the strength of the interactions between fluctuations, inversely proportional to the characteristic fluctuation interaction correlation length. \\ \\

The B20 fluctuation-induced first order transition was first observed in MnSi\cite{janoschek_fluctuation_2013, kindervater_critical_2014}, where the fluctuation interaction correlation length exceeds that of the DMI. The result is the Brazovskii scenario\cite{brazovskii_phase_1975}, where the diverging correlation length of the system crosses over first the DMI correlation length, then the fluctuation interaction correlation length. The result is that as these critical fluctuations begin to interact, they already have a chiral character. However, in the case of Cu$_2$OSeO$_3$, the fluctuation-induced transition follows a Wilson-Fisher scenario\cite{zivkovic_critical_2014, sabri_monte_2023}. This means that stronger interactions between fluctuations result in a correlation length that is too small for the appearance of chirality. While the fluctuations do not appear on a sphere in reciprocal space, as can be observed in MnSi\cite{janoschek_fluctuation_2013}, the existence of the first and second order transitions in close proximity can be corroborated by measurements of heat capacity\cite{adams_long-wavelength_2012} and magnetic entropy\cite{chauhan_multiple_2019}. \\ \\

Due to the diverging correlation length across a continuous phase transition, the free energy is typically dominated by a singular term that in turn gives rise to universal scaling behaviour. In magnetic systems, this results in the power-law relations: 
\begin{align}\label{eq: Beta equation}
    M \propto | T - T_C | ^{\beta}, \qquad T < T_C
\end{align}    
\begin{align}\label{eq: Gamma equation}
    \chi \propto \left[ T - T_C \right] ^{-\gamma}, \qquad T > T_C
\end{align} 
\begin{align}\label{eq: Delta equation}
    M\propto\left|H\left(T_C\right)\right|^\frac{1}{\delta_C} , \qquad T = T_C
\end{align} 
Where the magnetisation, $M$, is the order parameter of the transition, and $\chi$ is the magnetic susceptibility. $\delta$, $\beta$, and $\gamma$ are the exponents of the critical isotherm, magnetisation, and susceptibility  respectively. Due to the dependence of these exponents on only a small number of factors such as symmetry of the order parameter, the dimensionality of the system, and the spatial range of the interactions. The magnetisation critical exponent, $\beta$, and the susceptibility critical exponent, $\gamma$, then appear in the Arrott-Noakes relation\cite{arrott_approximate_1967}:
\begin{align}\label{eq: Arrott-Noakes Relation}
    \left[H/M\right]^{1/\gamma}=\left[T-T_C\right]/T_C+\left[M/M_1\right]^{1/\beta}
\end{align}   
Where $M_1$ is a material constant\cite{fan_critical_2010}. This describes the magnetic behaviour of a non mean-field magnetic system dominated by the SEI, and therefore applies to Cu$_2$OSe$_{1-x}$Te$_x$O$_3$ only at high magnetic field where the contributions of the DMI and magneto-crystalline anisotropy are low\cite{chauhan_different_2022, fan_critical_2010}.  Correct critical exponents will therefore yield plots of the relation $\left[H/M\right]^{1/\gamma}$ vs $M^{1/\beta}$ which display linear behaviours at high magnetic field, with constant gradient across the set of isotherms, denoted by $S(T)$. This means that the most accurate critical exponents $\left(\beta, \gamma\right)$ will be those that produce the set of normalised slopes $NS(T) = S(T)/S(T_C)$ which are closest to 1. \\ \\

This situation can be somewhat complicated in the presence of "dangerously irrelevant" anisotropies\cite{leonard_critical_2015}. These are higher order terms in the free energy, which would typically not be expected to contribute to the singular behaviour of the system, but nonetheless affect the scaling behaviour on one side of the transition to the point of changing the universality class. In the case of Cu$_2$OSeO$_3$, it is the cubic magnetic anisotropy that appears to modify the scaling behaviour across the PM to FD transition through a fourth order term in the free energy\cite{bak_theory_1980} . \\ \\

To describe the PM to FD phase transition in undoped Cu$_2$OSeO$_3$ using multiple universality classes, Chauhan \textit{et. al.}\cite{chauhan_different_2022} split the normalised slopes into separate datasets above and below the PM to FD critical temperature, $T_C$. This allowed the scaling behaviour to be described by separate exponents $\left(\beta_L, \gamma_L\right)$ and  $\left(\beta_H, \gamma_H\right)$, where the subscripts H and L refer to whether the scaling behaviour is above or below $T_C$. While the critical exponents could be found by interaction which brings the normalised slopes closest to 1, we find that there is, in principle, a line of suitable exponents in the $\left(\beta, \gamma\right)$ space\cite{Supplementary_Information} and as such, the results depend strongly on the starting parameters. We instead suggest the more robust method of calculating the normalised slopes above and below $T_C$, for each point in a reasonable space of critical exponents. The closeness of these slopes to 1, can then be quantified as the sum of square residuals. Due to the set of possible minima, the best exponents must be found using the constraint of the experimentally determined $\delta_C$ exponent. This can be related to the  $\left(\beta, \gamma\right)$ exponents through the Widom scaling relation\cite{widom_equation_1965}:
\begin{align}\label{eq: Widom scaling relation}
    \delta_C = \gamma/\beta+1
\end{align}
The selected critical exponents will then be those with the minimum normalised slope residual along the line $\gamma_{L/H} = \left[\delta_C-1\right]\beta_{L/H}$. The process of selecting the best pair of critical exponents from the  $\left(\beta, \gamma\right)$ space can be depicted on a heat-map of normalised slope residuals, which we refer to accordingly as the Heat-map Modified Iteration Method (HMIM). \\ \\

Using the original MIM, Chauhan \textit{et al.}\cite{chauhan_different_2022} showed that there is a transition in universality classes in undoped Cu$_2$OSeO$_3$, where the spin dimensionality reduces from Heisenberg-like to Ising-like as the temperature lowers past $T_C$. A similar result was found in the chiral magnetic soliton host Cr$_{1/3}$TaS$_2$\cite{meng_crossover_2023}, with the paramagnetic to chiral magnetic soliton transition also being described by the Heisenberg-like to Ising-like universality class transition . These transitions imply a significant change in anisotropy across the transition, with spins becoming more closely confined to an easy axis. \\ \\

The complex interplay of interactions that result in the range of magnetic phases, as well as this universality class transition and the fluctuation-induced first order transition in its vicinity, suggest that there are potentials to create bespoke materials for specific applications by tweaking the fundamental interaction strengths. Previous results using hydrostatic pressure show a widening of the stability range for the skyrmion lattice, as well as an increase in the skyrmion critical temperature\cite{levatic_dramatic_2016, wu_physical_2015}, which is predominantly attributed to an increase in MCA\cite{levatic_dramatic_2016}. Additionally, the opposite effect has been observed for Te-doping on the Se sites, resulting in lattice expansion of the crystal structure\cite{wu_physical_2015}. The expansion of the unit cell of Cu$_2$OSe$_{1-x}$Te$_x$O$_3$ via chemical doping with Te ($x$) lowers both the magnetic phase transition temperatures and the range of skyrmion thermodynamic stability. In our previous work, we showed via Lorentz Transmission Electron microscope of  Cu$_2$OSe$_{0.993}$Te$_{0.007}$O$_3$ that the skyrmion size and the helical modulation period is increased by Te-doping\cite{han_scaling_2020}, suggesting a change in the ratio $\lambda \propto J/D$ (see ref. \citen{kanazawa_noncentrosymmetric_2017}). This suggests that non-magnetic doping has the potential to affect both the Heisenberg to Ising transition through a modification of the MCA, and the FD phase through a modification of the DMI strength. 
\\ \\
In this work, we investigate this hypothesis by applying the novel HMIM to the PM to FD phase transition in single crystal and polycrystalline samples of Cu$_2$OSe$_{1-x}$Te$_x$O$_3$ $(0\leq x \leq 0.1)$. We used synchrotron powder X-ray diffraction to investigate the structural change upon doping. The helimagnetic orderings at specific magnetic fields and temperature points were confirmed with small-angle neutron scattering (SANS). We show that the Heisenberg to Ising transition is robust to Te-doping. Finally we use phenomenological arguments to quantify the changes in the strength of the DMI, SEI, and MCA.  \\ \\

\section*{Results and Discussion}

High-resolution synchrotron powder X-ray diffraction (pXRD) patterns for the polycrystalline Cu$_2$OSe$_{1-x}$Te$_x$O$_3$ samples are included in Fig. S1\cite{Supplementary_Information} to show the phase purity and structure of the samples synthesised. Upon successfully doping Te ions into the Se site, the unit cell expands to accommodate for the larger ion being present, which shifts the diffraction peaks to lower 2$\theta$ values. When x > 0.1, samples contained a small amount of an impurity phase, Cu$_3$TeO$_6$. As such, the current analysis uses samples up to x $\leq$ 0.1 to minimise the effects of the impurity phases. The Rietveld refinement analysis indicates that the actual average doping of the samples was 5.32 \% for the nominal Cu$_2$OSe$_{0.95}$Te$_{0.05}$O$_3$ sample and 8.01 \% for the nominal Cu$_2$OSe$_{0.9}$Te$_{0.1}$O$_3$ sample. The higher Te-doping for the nominal Cu$_2$OSe$_{0.95}$Te$_{0.05}$O$_3$ sample most likely arises from the high volatility of SeO$_2$, yielding a higher Te/Se ratio present in the powder. The nominal Cu$_2$OSe$_{0.9}$Te$_{0.1}$O$_3$ sample has a lower amount of Te-doping present, 8.01 \%, as the Te starts to be included in the Cu$_3$TeO$_6$ impurity phase.\\ \\

\begin{figure}[ht]
\centering
\includegraphics[width=\linewidth]{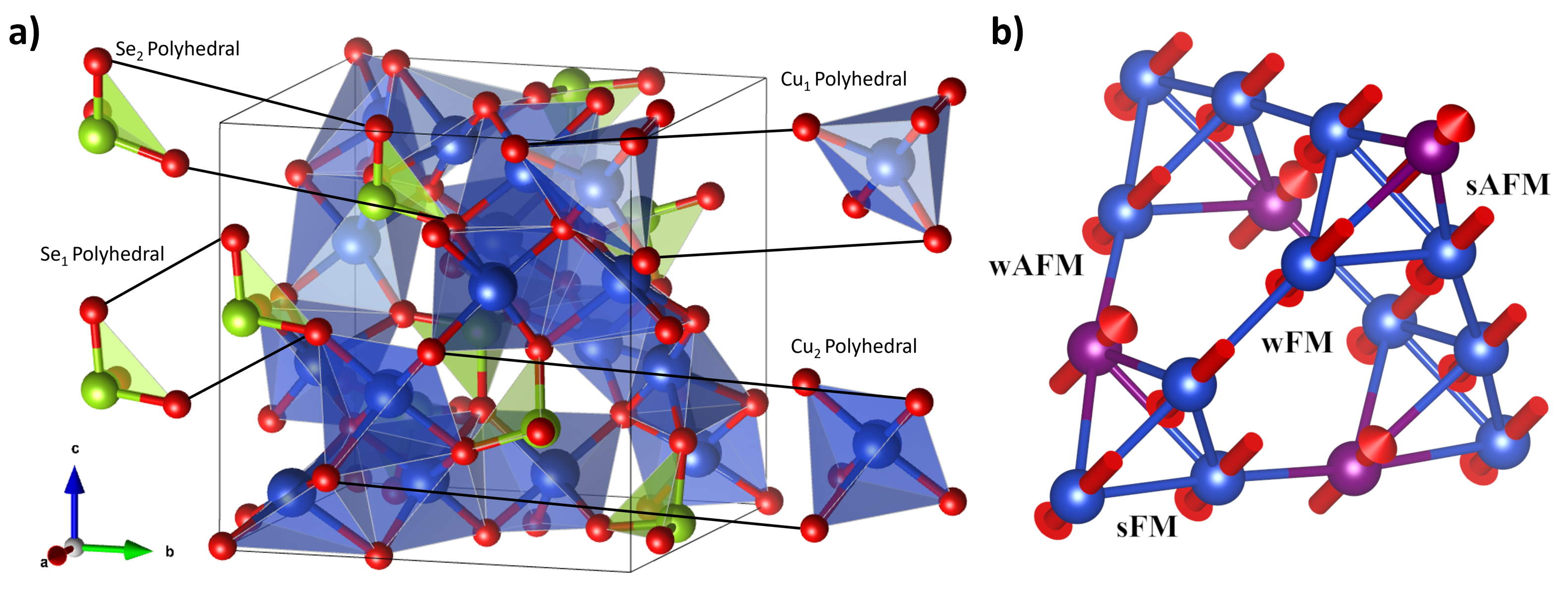}
\caption{a) The unit cell for Cu$_2$OSeO$_3$ modelled in VESTA 3. \cite{Momma:db5098} Blue atoms are Copper, lime atoms are Selenium, and red atoms are Oxygen. The cutout shows the two different Cu and Se polyhedra. b) The structure of the magnetic ions with labelled Cu-Cu interactions; these are not 'physical' Cu-Cu bonds and aren't shown in the unit cell figure. The magnetic moments are shown with red arrows in their typical 3-down-1-up arrangement. The Cu$_1$ site is depicted in purple, while the Cu$_2$ site is blue. The interaction lengths are categorized as strong and weak ferromagnetic (sFM and wFM) as well as strong and weak antiferromagnetic (sAFM and wAFM), as described in Janson \textit{et. al.}\cite{janson_quantum_2014}} 
\label{fig:Unit Cell Info}
\end{figure}

The crystallographic unit cell for Cu$_2$OSeO$_3$, is shown in part a of Fig. \ref{fig:Unit Cell Info}. In the unit cell, there are two CuO$_5$ polyhedra for the two different Cu sites along with the two SeO$_3$ polyhedra. The Cu$_1$ site has a trigonal bipyramidal coordination, while the Cu$_2$ site has a square pyramidal coordination with oxygen atoms, while both Se sites have a trigonal pyramidal coordination. Cu$_2$OSeO$_3$ crystallises in a cubic space group, \textit{P2$_1$3}, resulting in the lattice constants, \textit{a} = \textit{b} = \textit{c}, thus the lattice constant, \textit{a}, can be plotted against the Te-doping (x). This shows a linear trend with the expansion of the unit cell with the increasing incorporation of Te into the structure as shown in Fig. S1\cite{Supplementary_Information}. \\ \\

The magnetic interactions rely on the spatial arrangement of the Cu$^{2+}$ ions in the crystal structure; thus, examining the changes in bond interaction lengths with Te-doping is important. Upon doping Te into the Se sites, the unit cell expands, resulting in the overall lengthening of the Cu-Cu interaction lengths as presented n Table \ref{tab:Synchrotron Interaction Lengths}. These values are derived from synchrotron pXRD data, which has a higher spatial resolution from the shorter wavelength used and better accuracy due to a higher signal-to-noise ratio compared to other published value for Te-doped samples\cite{wu_physical_2015}. The labels chosen to differentiate the Cu-Cu interaction lengths are based on the type of magnetic interaction, either ferromagnetic (FM) between spin-aligned Cu sites or antiferromagnetic (AFM) for opposing Cu site spins, as shown in part b of Fig. \ref{fig:Unit Cell Info}. For the strong FM (sFM) interaction, there is better electronic density overlap despite the lattice expansion, unlike the AFM interaction, where the greater bond length has a weaker overlap. There is a clear increase in bond length from the undoped to the nominal 10 \% Te-doped sample for the wAFM, wFM and sFM, while this is not the case for the sAFM. This could arise from the preference of Te-doping into the second Se site over the first Se site. 
\\ \\
\begin{table}[ht]
\centering
\renewcommand{\arraystretch}{1.5}
\begin{tabular}{|c|cccc|cc|}
\hline
\multirow{2}{*}{Nominal Composition (x)} & \multicolumn{4}{c|}{Cu-Cu Interaction Lengths (\AA)} & \multicolumn{2}{c|}{Lattice Parameters} \\ \cline{2-7} 
\multicolumn{1}{|c|}{}  & \multicolumn{1}{c|}{sAFM Cu$_1$-Cu$_2$} & \multicolumn{1}{c|}{wAFM Cu$_1$-Cu$_2$} & \multicolumn{1}{c|}{wFM Cu$_2$-Cu$_2$} & \multicolumn{1}{c|}{sFM Cu$_2$-Cu$_2$} & \multicolumn{1}{c|}{\textit{a} (\AA)} & {Volume (\AA$^{3}$)}\\ \hline
0 & \multicolumn{1}{c|}{3.0474(10)} & \multicolumn{1}{c|}{3.3098(10)} & \multicolumn{1}{c|}{3.0530(9)} & \multicolumn{1}{c|}{3.2229(10)} &\multicolumn{1}{c|}{8.92359(1)} & 710.589(1)\\ \hline
0.05 & \multicolumn{1}{c|}{3.0481(8)} & \multicolumn{1}{c|}{3.3119(8)} & \multicolumn{1}{c|}{3.0555(8)} & \multicolumn{1}{c|}{3.2228(8)} & \multicolumn{1}{c|}{8.92743(1)} & 711.506(1)\\ \hline
0.1 & \multicolumn{1}{c|}{3.0469(8)} & \multicolumn{1}{c|}{3.3158(8)} & \multicolumn{1}{c|}{3.0573(8)} & \multicolumn{1}{c|}{3.2229(8)} & \multicolumn{1}{c|}{8.93055(1)} & 712.253(1)\\ \hline 
\end{tabular}
\caption{Cu-Cu interaction lengths, lattice parameters and volumes of the unit cell for polycrystalline Cu$_2$OSe$_{1-x}$Te$_x$O$_3$ samples as refined from synchrotron pXRD data. Cu$_1$-Cu$_2$ and Cu$_2$-Cu$_2$ interaction lengths are labelled with strong and weak ferromagnetic (sFM and wFM), and strong and weak antiferromagnetic (sAFM and wAFM).}
\label{tab:Synchrotron Interaction Lengths}
\end{table}

A compositional analysis was also carried out on the polycrystalline samples using Energy Dispersive X-ray Spectroscopy (EDS) with the spectra shown in Fig S3-S5\cite{Supplementary_Information}. The results of this analysis are compared to those derived from Rietveld refinement. The nominal Cu$_2$OSe$_{0.95}$Te$_{0.05}$O$_3$ sample has a value of x = 5.32 \% and 4.46 \%, while the nominal Cu$_2$OSe$_{0.9}$Te$_{0.1}$O$_3$ sample has a value of x = 8.00 \% and 6.34 \%, from Rietveld refinement and EDS respectively. The slight variation in values between both techniques can be understood from the difference in the characterisation techniques. In particular, EDS can only differentiate elements but not phase. For the two Te-doped samples, there are minimal impurities (< 3 \%), thus it is assumed that the atomic percentages of Te present is what is incorporated into the Cu$_2$OSeO$_3$ structure. Irrespective of their values, both measurements demonstrate that the two samples have increasing concentrations of Te doped into the crystal structure.
\\ \\
The magnetic phases that are present in this B20 material are distinguishable via unique magnetic SANS patterns. Detector images of the helical, conical and skyrmion phases are established for both single crystal\cite{muhlbauer_skyrmion_2009, baral2022tuning} and polycrystalline\cite{sukhanov2019increasing} samples of Cu$_2$OSeO$_3$, allowing these phases to be checked for their robustness to Te-doping. For polycrystalline samples when below $T_C$, the material is a helimagnet, as the SANS detector image is represented by a uniform ring pattern with no distinguishable peaks \cite{sukhanov2019increasing}. For single crystals, the helical phase is observed as two Bragg peaks separated by 180 degrees \cite{chacon2018observation}. Above a critical magnetic field, when a field is applied, each helical domain of the individual crystallites aligns along the direction of the field to form the conical phase. This is represented in the SANS detector images as two intense Bragg peaks at the same q$_y$ position horizontally in the detector plane or perpendicular to the incident neutron beam. In the conditions where the skyrmion phase is formed, the feature is distinguished by two peaks perpendicular to the conical peaks in the vertical direction of the detector plane. The simultaneous observation of both skyrmion and conical phases in polycrystalline samples is unlike SANS measurements of single crystals since the projection planes of the conical and skyrmion phases are orthogonal to each other \cite{chacon2018observation}.As such, the conical phase is not visible in the standard experimental configuration to observe the skyrmion phase where the magnetic field is parallel with the neutron beam  \\

\begin{figure}[h]
\centering
\includegraphics[width=\linewidth]{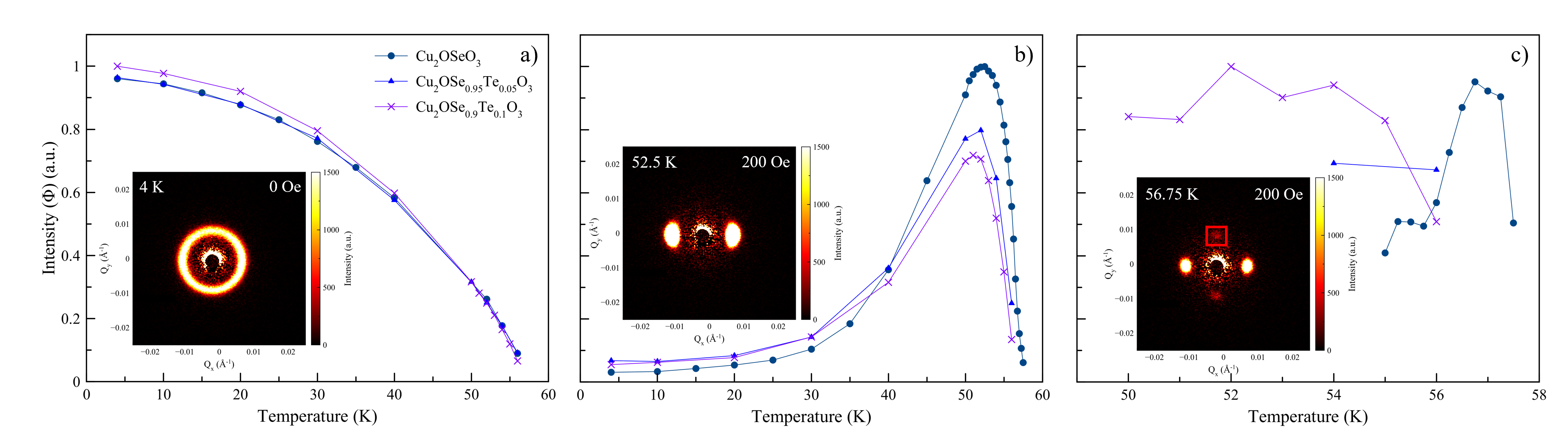}
\caption{The normalised SANS peak intensities as a function of temperature for polycrystalline samples of Cu$_2$OSeO$_3$, Cu$_2$OSe$_{0.95}$Te$_{0.05}$O$_3$ and Cu$_2$OSe$_{0.9}$Te$_{0.1}$O$_3$. a) Radial integrated peak intensity with zero applied magnetic field extracted from the SANS patterns, corresponding to the helical phase. The inset shows the detector image of the helical pattern at the maximum intensity for Cu$_2$OSeO$_3$ at 4 K. b) Annular peak intensity with 200 Oe applied magnetic field perpendicular to the neutron beam, corresponding to the conical phase. The inset shows the detector image of the conical pattern for the maximum intensity for Cu$_2$OSeO$_3$ at 52.5 K. c) Annular peak intensity with 200 Oe applied magnetic field, corresponding to the skyrmion phase. The inset shows the detector image of the conical pattern and the skyrmion pattern at the maximum intensity for Cu$_2$OSeO$_3$ at 56.75 K with the red box to guide the eye to the skyrmion pattern.}
\label{SANS Figure}
\end{figure}

As shown in Fig. \ref{SANS Figure}a), all polycrystalline samples exhibit helimagnetic ordering at 0 Oe. The uniform ring pattern characterises the helimagnetic phase in polycrystalline samples\cite{sukhanov2019increasing} due to the random orientation of the crystalites and the intensity increasing as the temperature decreases. The SANS detector images in Fig. S6 \cite{Supplementary_Information} shows that both the Cu$_2$OSe$_{0.95}$Te$_{0.05}$O$_3$ and Cu$_2$OSe$_{0.91}$Te$_{0.1}$O$_3$ samples retain its helimagnetic ordering; this is in agreements with Wu et al. \cite{wu_physical_2015}. \\ \\ 

Fig. \ref{SANS Figure}b) shows the conical phase, characterised by two q$_y=0$ peaks at the same $|\mathrm{q}|$\cite{sukhanov2019increasing}. At 200 Oe, the phase is suppressed with increasing Te-doping, along with a shift to lower temperature for the intensity maximum from 52.5 K in the undoped sample to 51 K in the Cu$_2$OSe$_{0.9}$Te$_{0.1}$O$_3$ sample. The intensity of the conical patterns increases with temperature as the systems shift out of the helical phase, which can be seen in the Fig. 7 \cite{Supplementary_Information}. Additional skyrmion patterns for all samples are presented, as well as FD regime patterns consistent with the Wilson-Fisher scenario, suggesting that the strength of the DMI has, at a minimum, not significantly increased relative to the fluctuation interaction strength.     \\ \\

In Fig. \ref{SANS Figure}c) it can be clearly observed that the skyrmion phase,  shifts to lower temperatures with higher Te-doping. Only two temperature points were taken in the skyrmion region for the Cu$_2$OSe$_{0.95}$Te$_{0.05}$O$_3$ sample, but this still shows a visible SANS skyrmion pattern at lower temperatures than the undoped sample. The Cu$_2$OSe$_{0.9}$Te$_{0.1}$O$_3$ sample has a much larger temperature range of skyrmion stability, from 50 - 56 K as compared to 55 - 57.5 K in the undoped sample, in contrast with the magnetic susceptibility results of Wu \textit{et al.}\cite{wu_physical_2015}. This is likely due to the coexistence of the skyrmion and conical phases, as highlighted in Fig. \ref{SANS Figure}c). Depending on the field and temperature conditions, the conical phase has a greater detector intensity compared to the skyrmion phase which can make it more difficult to observe the skyrmion peaks without intensity scaling. This paints a different picture compared to single crystal SANS measurements which due to the orthogonal relationship in the projection of skyrmion and conical phase, only one of the magnetic phases can be observed in each scattering configuration \cite{chacon2018observation} thus concealing the full story of both magnetic phases' coexistence. \\ \\

\begin{figure}[ht]
\centering
\includegraphics[width=\linewidth]{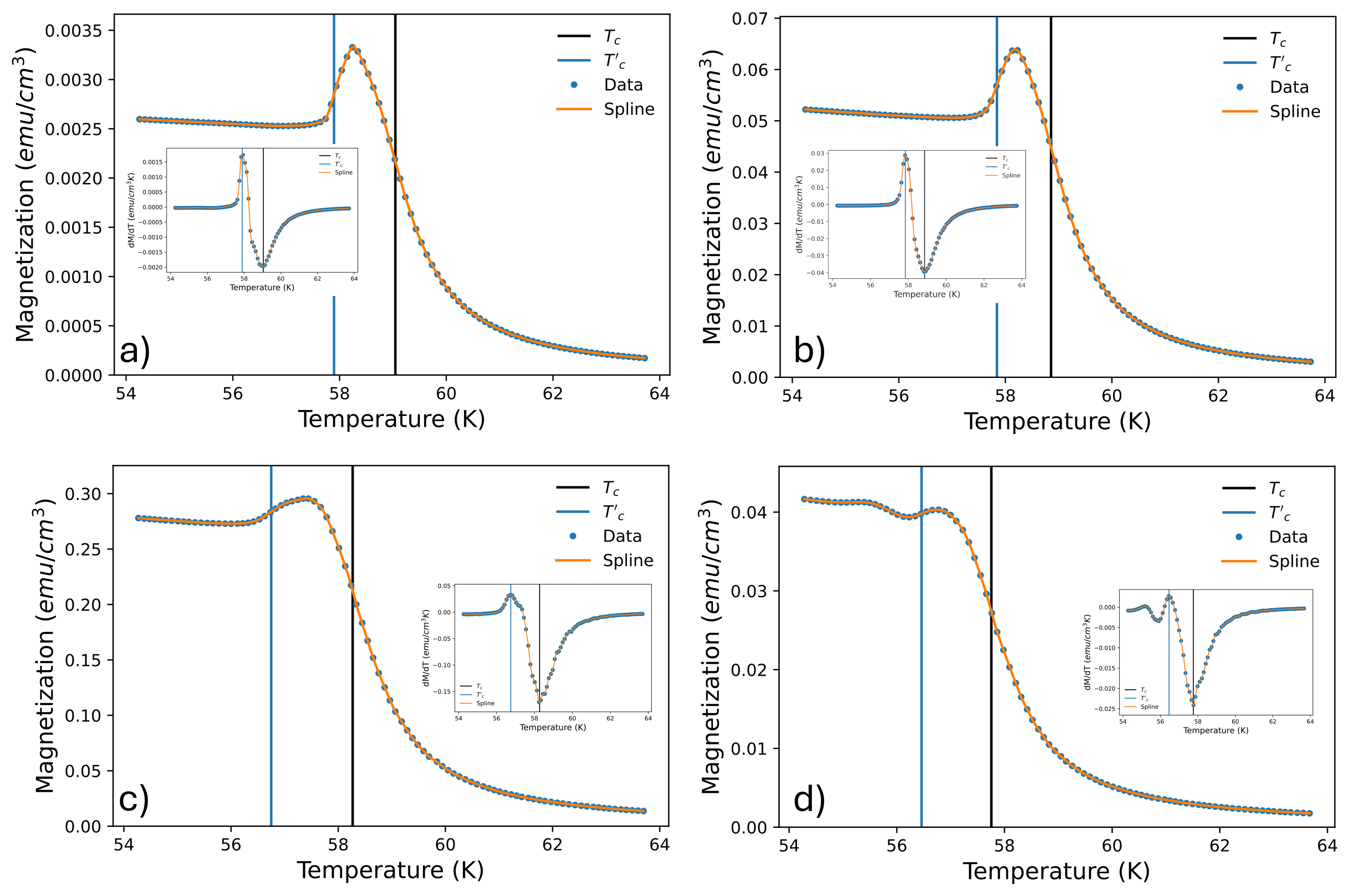}
\caption{The MvT measurements at zero magnetic field. a)-d) correspond to the data from the Cu$_2$OSeO$_3$ single crystal, Cu$_2$OSeO$_3$ pellet, Cu$_2$OSe$_{0.95}$Te$_{0.05}$O$_3$ pellet, and Cu$_2$OSe$_{0.9}$Te$_{0.1}$O$_3$ pellet, respectively. Insets: $dM/dT$ of the spline fit, used to determine the inflection points corresponding to $T_C$ and  $T'_C$ for the respective samples. It is important to note that the magnitude of this ‘zero field’ measurement is likely to be sensitive to both the remnant field in the MPMS 3 external magnet and the small measurement field of the SQUID. Every tenth data point is shown for visual clarity.} 
\label{fig:MvT}
\end{figure}

To obtain further details on the magnetic interactions within the material, magnetic measurements were performed using a Superconducting Quantum Interference Device (SQUID). The polycrystalline samples were pressed into pellets to minimise the effects of an internal field and hence allowing for a more accurate account of the demagnetisation field (SI). An unoriented undoped single crystal was also used to show the minimal effect pellet pressing has on the magnetic characteristic of the sample.  The critical temperatures corresponding to the PM to FD phase transition, $T_C$, and FD to HM phase transition, $T'_C$, were found through the inflection points in the magnetisation vs. temperature measurements at zero magnetic field\cite{chauhan_different_2022}. This is shown in Fig. \ref{fig:MvT}, where the first derivatives of magnetisation with respect to temperature are shown in the insets. The magnetisation-temperature data are fitted with a cubic spline. The results are shown in table \ref{tab:critical_exponents_1}. \\ \\

\begin{figure}[ht]
\centering
\includegraphics[width=\linewidth]{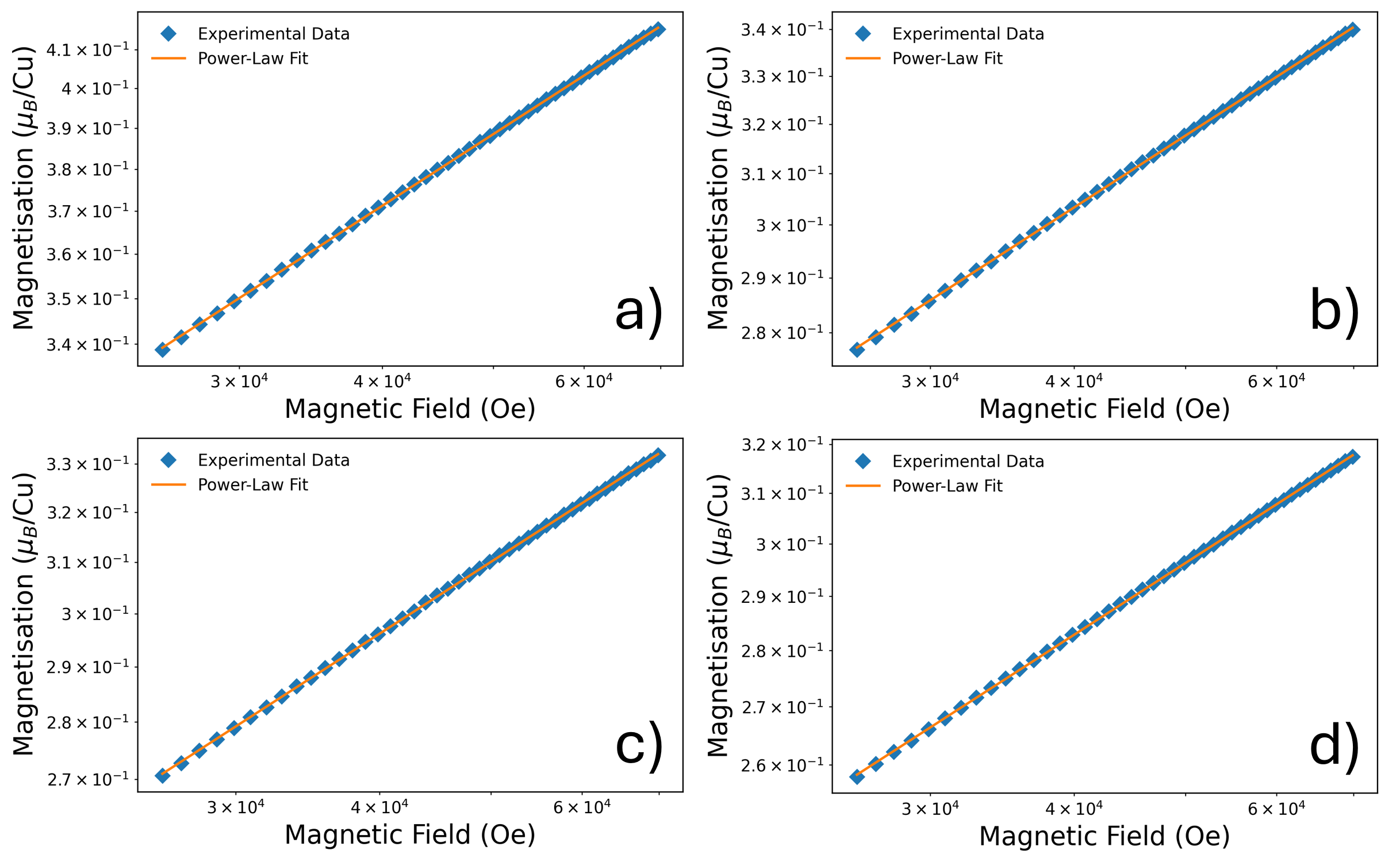}
\caption{Fits of the high field region of the critical isotherms on log-log plots. a)-d) are the Cu$_2$OSeO$_3$ single crystal, Cu$_2$OSeO$_3$ pellet, Cu$_2$OSe$_{0.95}$Te$_{0.05}$O$_3$ pellet, and Cu$_2$OSe$_{0.9}$Te$_{0.1}$O$_3$ pellet, respectively. The errorbars are smaller than the size of the markers.}
\label{fig:delta_fits}
\end{figure}

For the universality class analysis, $T_C$  was taken to be the nearest 0.5 K, such that the critical isotherms are 59 K, 59 K, 58 K, and 57.5 K for the undoped single crystal, and the Cu$_2$OSeO$_3$, Cu$_2$OSe$_{0.95}$Te$_{0.05}$O$_3$, and Cu$_2$OSe$_{0.9}$Te$_{0.1}$O$_3$ polycrystalline samples respectively. These isotherms are included in both the above and below critical temperature normalised slopes plots. The HMIM was applied to all three polycrystalline samples as well as the undoped single crystal. The first step is to fit a power-law function of the form $M\propto\left|H\left(T_C\right)\right|^\frac{1}{\delta_C}$ to the critical isotherm, allowing the critical delta exponent to be extracted from the slopes of the log-log plots. These are shown in Fig. \ref{fig:delta_fits}, and all have an $R^2>0.999$. The resulting $\delta_C$ are shown in table \ref{tab:critical_exponents_1} and range from 4.80 To 4.90. Their distribution corresponds to the theoretically expected $\delta_C$ exponents of the Ising, XY, and Heisenberg models, but not that of the tricritial mean-field model\cite{kim_critical_2002}. It is likely that the variation in critical delta seen across our samples is due to the accuracy of our data to the true critical temperature of the system, as changing the input $T_C$ value by $\pm0.5$ K typically resulted in changes in delta of $\pm0.2$. This was found to be insignificant for the Heisenberg to Ising transition for all samples, and did not change any of the $\left(\beta,\gamma\right)$ critical exponents materially.\\ \\

\begin{figure}[ht]
\centering
\includegraphics[width=\linewidth]{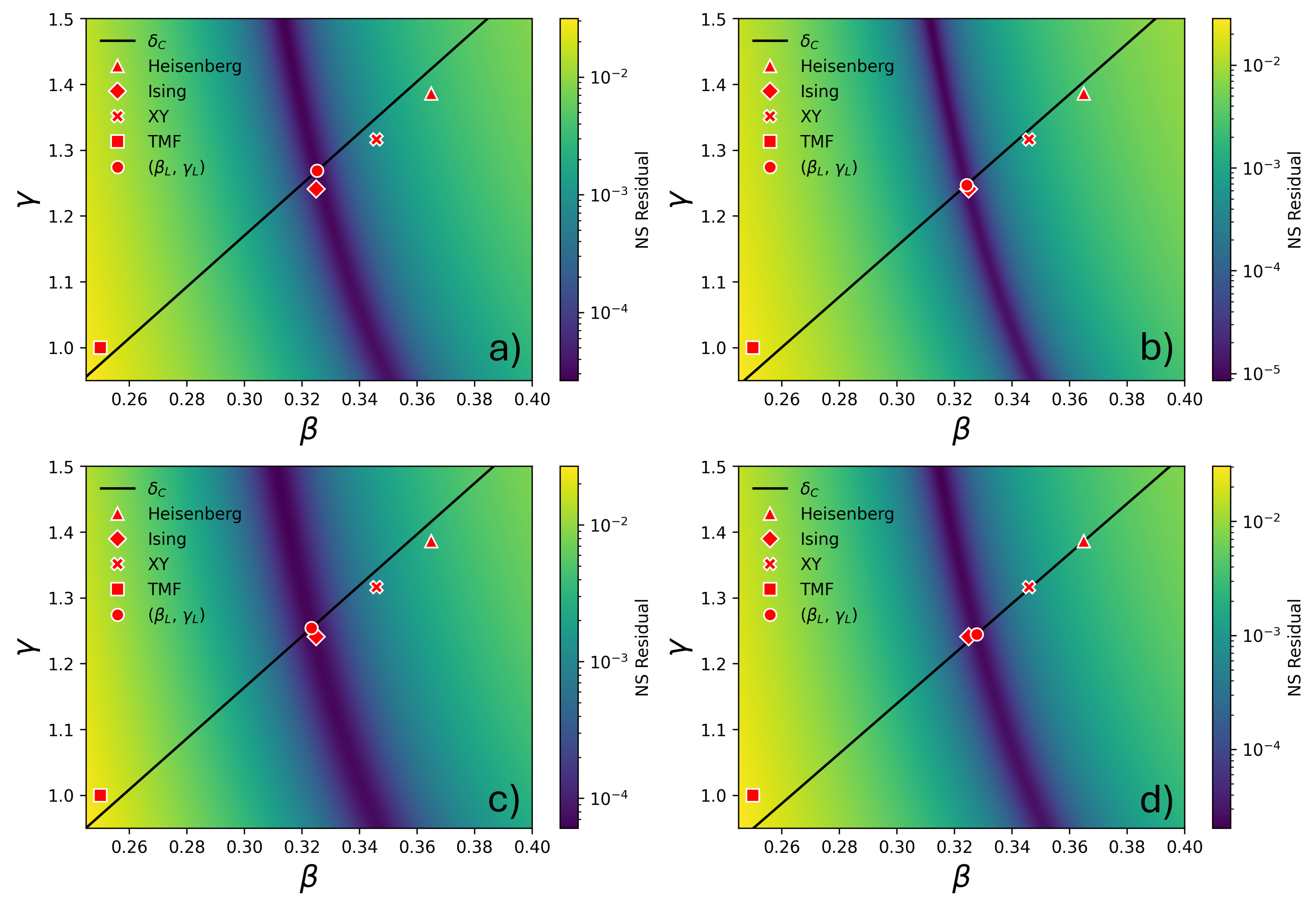}
\caption{The below $T_C$ normalised slope residuals plotted across the ($\beta$,$\gamma$) parameter space for each sample. a)-d) are the undoped single crystal, undoped pellet, 5\% Te-doped pellet, and 10\% Te-doped pellet respectively. The errorbars are smaller than the size of the markers.}
\label{fig:low_t_results}
\end{figure}

\begin{figure}[ht]
\centering
\includegraphics[width=\linewidth]{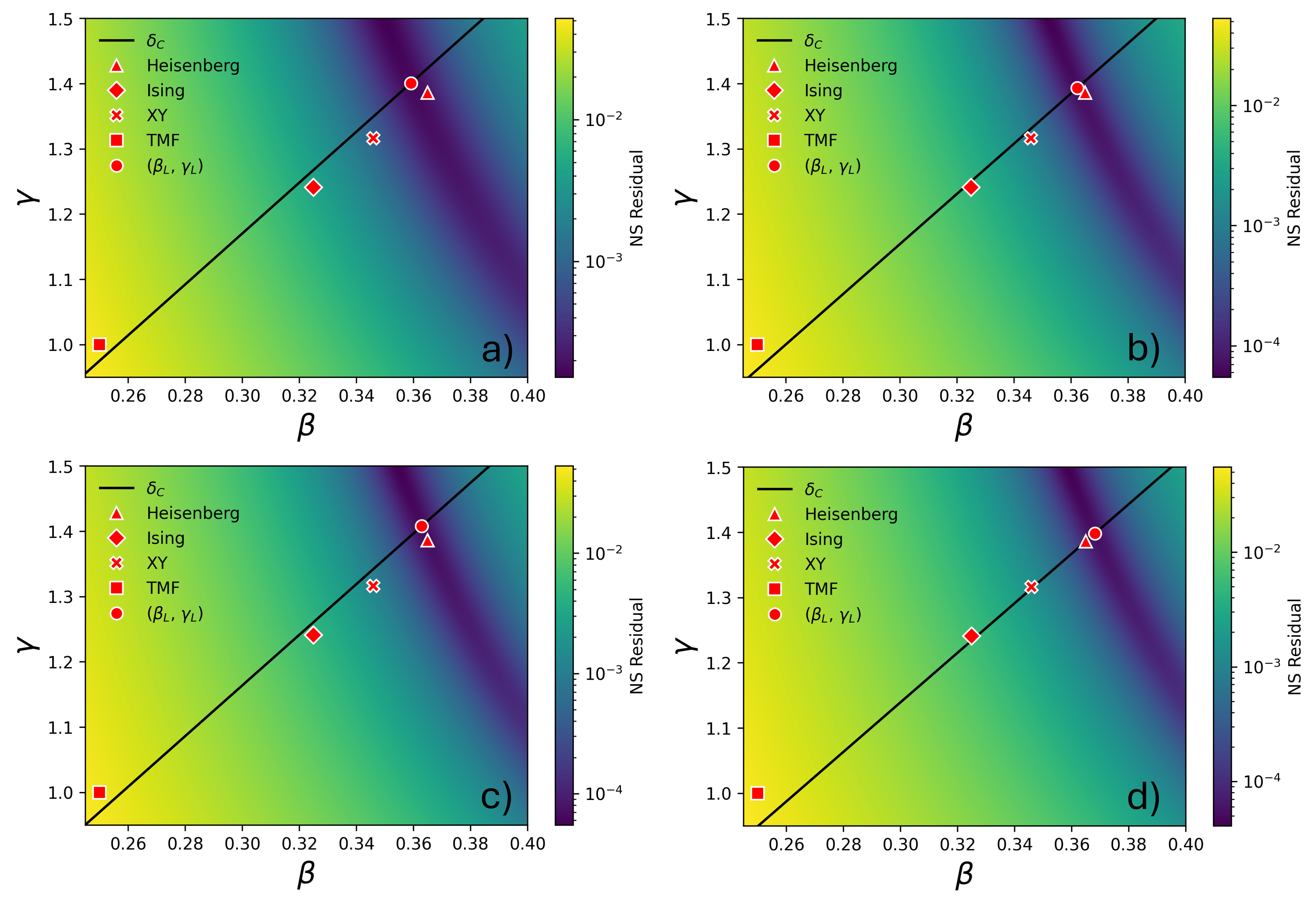}
\caption{The above $T_C$ normalised slope residuals plotted across the ($\beta$,$\gamma$) parameter space for each sample. a)-d) are the Cu$_2$OSeO$_3$ single crystal, Cu$_2$OSeO$_3$ pellet, Cu$_2$OSe$_{0.95}$Te$_{0.05}$O$_3$ pellet, and Cu$_2$OSe$_{0.9}$Te$_{0.1}$O$_3$ pellet respectively. The errorbars are smaller than the size of the markers.}
\label{fig:high_t_results}
\end{figure}

Shown in Fig. \ref{fig:low_t_results} are the heat-maps below $T_C$ for all 4 samples, with the common magnetic universality classes depicted. The ($\beta_L$,$\gamma_L$) found through this method are nearest to the Ising universality class. Additionally, the set of minima with roughly equal normalised slope residuals is shown for all samples as well as the above $T_C$ normalised slope residuals shown in Fig. \ref{fig:high_t_results}. The width of the set of normalised slope residual minima appears to have increased, which is likely to be due to the concavity exhibited by the normalised slope data in the paramagnetic phase. This is shown across the set of samples in Fig. \ref{fig:Normalised_Slopes}, through the normalised slopes of the selected ($\beta_L$,$\gamma_L$) and ($\beta_H$,$\gamma_H$) exponents.\\ \\

The ($\beta$,$\gamma$) results from the above and below $T_C$ HMIM analysis are summarised in Fig. \ref{fig:Heatmap_MIM_results}. In Fig. \ref{fig:Heatmap_MIM_results}a), the below $T_C$ critical exponents are clustered around the Ising universality class at $(0.325, 1.241)$ (see ref. \citen{kim_critical_2002}) in parameter space, while in the above $T_C$ plot in Fig. \ref{fig:Heatmap_MIM_results}b) the results are clustered around the Heisenberg universality class at $(0.365,1.386)$ (see ref. \citen{kim_critical_2002}). The insets depict the results with the 95\% confidence intervals as derived from the areas in parameter space of the normalised slope residual minima. While there is a small deviation from the respective universality classes for  samples, this can arise due to the method of error analysis not considering the variability of results due to the rounding of critical temperatures. \\ \\

\begin{figure}[ht]
\centering
\includegraphics[width=\linewidth]{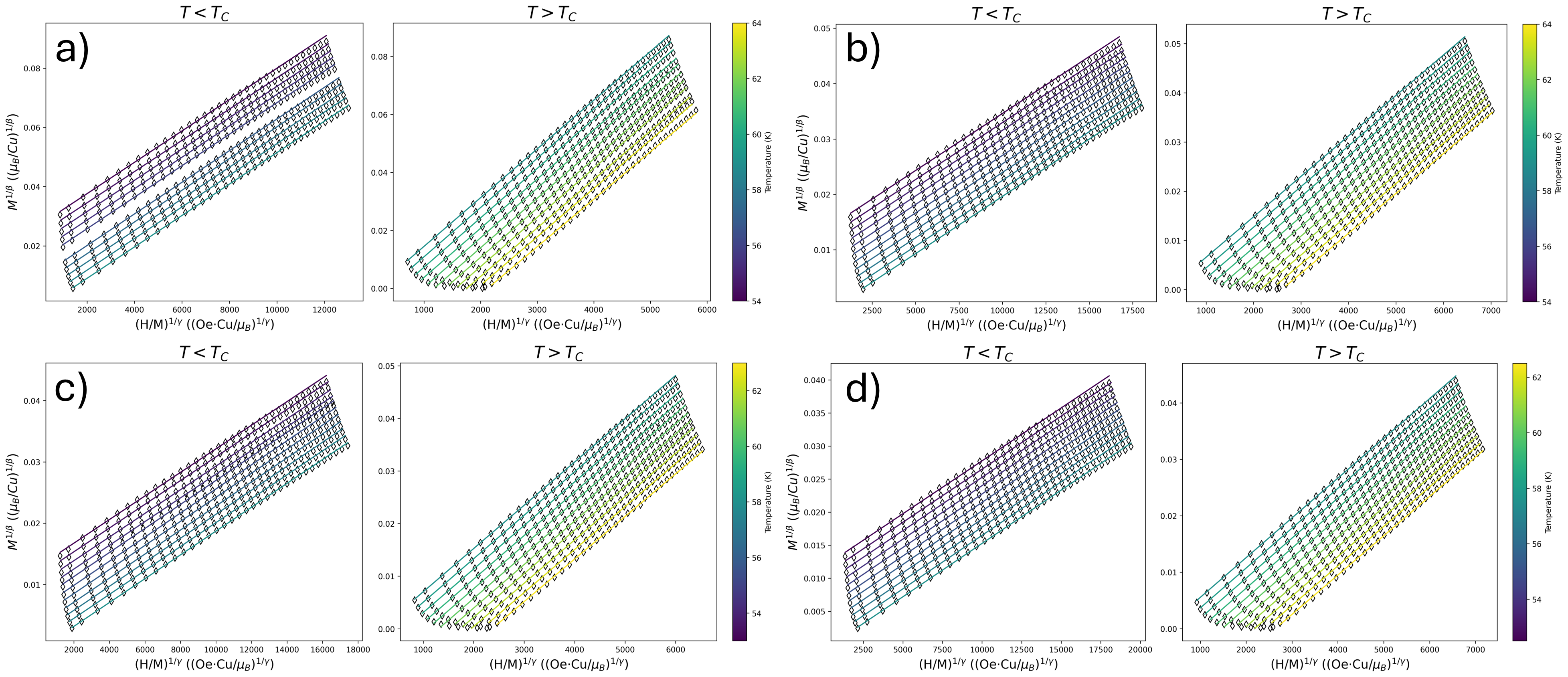}
\caption{The modified Arrott plots corresponding to the exponents found through HMIM. Only every second datapoint in each isotherm is shown for visual clarity. a)-d) are the Cu$_2$OSeO$_3$ single crystal, Cu$_2$OSeO$_3$ pellet, Cu$_2$OSe$_{0.95}$Te$_{0.05}$O$_3$ pellet, and Cu$_2$OSe$_{0.9}$Te$_{0.1}$O$_3$ pellet respectively. Note: a temperature line is removed in a) due to an outlier in the normalised slope.}
\label{fig:MAPs}
\end{figure}

\begin{figure}[ht]
\centering
\includegraphics[width=\linewidth]{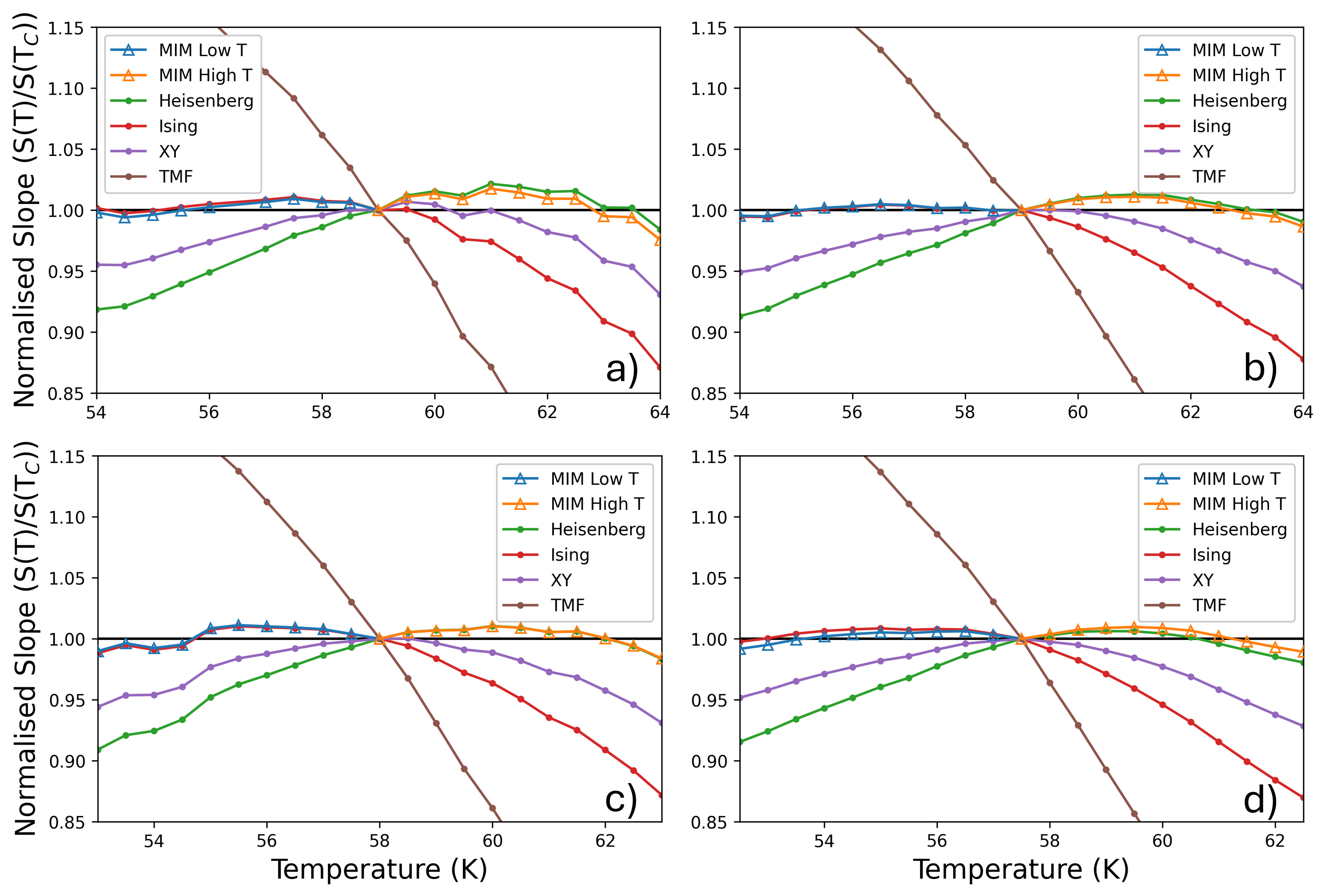}
\caption{Normalised slope plots for the selected exponents. a)-d) are the Cu$_2$OSeO$_3$ single crystal, Cu$_2$OSeO$_3$ pellet, Cu$_2$OSe$_{0.95}$Te$_{0.05}$O$_3$ pellet, and Cu$_2$OSe$_{0.9}$Te$_{0.1}$O$_3$ pellet respectively. Common magnetic universality classes are depicted, including the Heisenberg, Ising, XY, and Tricritical Mean-Field (TMF). Critical exponents for these universality classes are listed in ref. [\citen{chauhan_different_2022}].}
\label{fig:Normalised_Slopes}
\end{figure}

\begin{figure}[ht]
\centering
\includegraphics[width=\linewidth]{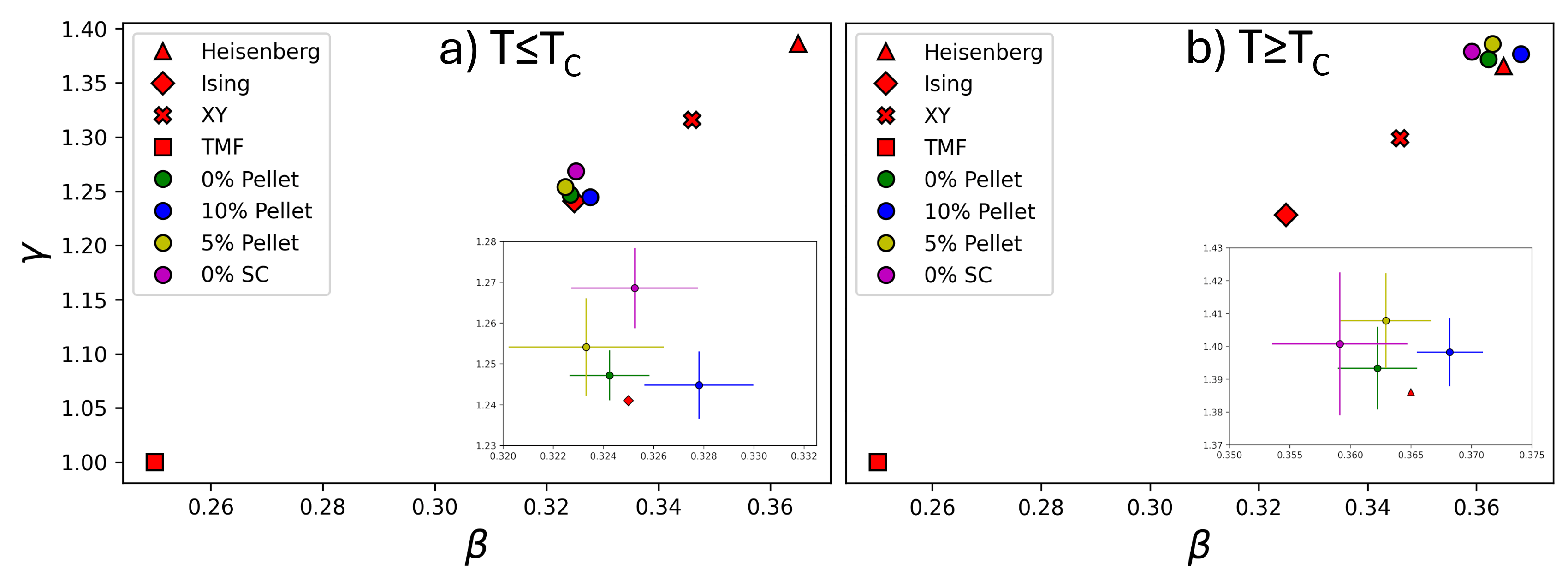}
\caption{The results of the HMIM for the data taken a) below $T_C$, and b) above $T_C$, alongside relevant magnetic universality classes. The errorbars are too small to be visible on this plot. Insets: closer view of the critical exponents, with 95\% confidence intervals calculated through the MAP fitting errors.}
\label{fig:Heatmap_MIM_results}
\end{figure}

These high and low exponents result from the modified Arrott plots (MAPs) shown in Fig. \ref{fig:MAPs}. Fig. \ref{fig:MAPs}a) shows the MAPs of the single crystal samples. The normalised slopes of the modified Arrott plots in Fig. \ref{fig:MAPs} are shown in Fig. \ref{fig:Normalised_Slopes}. As shown in Fig. \ref{fig:Normalised_Slopes}a), the undoped single crystal data appears to be significantly noisier that the others, particularly above $T_C$. This is likely due to the lower sample mass of the single crystal as compared with the pellets. It is also worth noting that all samples appear to show a slight concavity in their normalised slopes. This is commonly observed in normalised slope plots\cite{zhang_critical_2016,meng_crossover_2023}, and may be due to the system moving away from temperatures at which the coherence length diverges, resulting in non-universal behaviour. \\\\

Further information about the magnetic interactions of the samples can be extracted through the calculation of the exchange distance constant $\sigma$, where $J(r)\propto r^{-(d+\sigma)}$ with d=3 being the spatial dimensionality. This can be calculated by solving the quadratic equation (\ref{gamma_quadratic})\cite{chauhan_different_2022}: 
\begin{equation}\label{gamma_quadratic}
    \gamma=1+\frac{4(n+2)}{d(n+8)}\left[\sigma-d/2\right]+\frac{8(n+2)(n-4)}{d^2(n+8)^2}\left[1+\frac{2(7n+20)(3-d^2/8)}{(n-4)(n+8)}\right]\left[\sigma-d/2\right]^2
\end{equation} \\

The resulting critical exponents are listed in table (\ref{tab:critical_exponents_1}). This includes not just the $\beta$,$\gamma$, $\delta$, and $\sigma$ exponents, but also the $\alpha$ and $\nu$ exponents. The terms $\alpha$ and $\nu$ relate to the specific heat capacity through $C\propto\tau^{-\alpha}$, and the correlation length through $\xi\propto\tau^{-\nu}$, respectively. These can be calculated through the scaling relations $\nu=\frac{\gamma}{\sigma}$, and $\alpha=2-2\beta-\gamma$. The $\alpha$ exponents appear to agree with the Heisenberg to Ising transition. The undoped samples reach statistical agreement with the Heisenberg exponent of $\alpha=-0.115$\cite{kim_critical_2002} above $T_C$, while the Te-doped samples are narrowly lower than the Heisenberg value above $T_C$. The $\alpha$ exponents below $T_C$ agree with the Ising $\alpha=0.11$, with the exception of the undoped single crystal which has a slightly lower value of $\alpha$.\\ \\
\begin{table}[ht]
\centering
\begin{tabular}{|l|c|c|c|c|c|c|c|}
\hline
Sample  & $T_C$  &  $T'_C$  &$\beta_L$ & $\gamma_L$ & $\beta_H$ & $\gamma_H$ & $\delta_C$  \\ \hline
Cu$_2$OSeO$_3$ Single Crystal & 59.05 & 57.90 & 0.3253(25) & 1.2686(98) & 0.3591(55) & 1.401(21) & 4.9000(2)    \\ \hline
Cu$_2$OSeO$_3$ Pellet  & 58.86 & 57.84 & 0.3242(16) & 1.2472(61)  & 0.3622(32) & 1.393(13) & 4.8464(3)    \\ \hline
Cu$_2$OSe$_{0.95}$Te$_{0.05}$O$_3$ Pellet & 58.24 & 56.73 & 0.3233(30) & 1.254(12) & 0.3629(27) & 1.408(14) & 4.8790(3)     \\ \hline
Cu$_2$OSe$_{0.9}$Te$_{0.1}$O$_3$ Pellet & 57.74 & 56.46 &  0.3278(22)  & 1.2448(83) & 0.3682(37) & 1.399(10) & 4.7972(3)   \\ \hline        
\end{tabular}
\end{table}

\begin{table}[ht]
\centering
\begin{tabular}{|l|c|c|c|c|c|c|}
\hline
Sample & $\sigma_L$ & $\sigma_H$ & $\alpha_L$ & $\alpha_H$ & $\nu_L$ & $\nu_H$ \\ \hline
Cu$_2$OSeO$_3$ Single Crystal & 1.939(12) & 1.950(19) & 0.0809(111) & -0.1190(245) & 0.6544(66) & 0.7182(130) \\ \hline
Cu$_2$OSeO$_3$ Pellet & 1.911(8) & 1.944(11) & 0.1043(69) & -0.1180(142) & 0.6527(42)  & 0.7168(76) \\ \hline
Cu$_2$OSe$_{0.95}$Te$_{0.05}$O$_3$ Pellet & 1.920(16) & 1.956(12) & 0.0993(135) & -0.1337(163) & 0.6532(82) & 0.7196(87)  \\ \hline
Cu$_2$OSe$_{0.9}$Te$_{0.1}$O$_3$ Pellet & 1.908(11)  & 1.948(9) & 0.0996(93) & -0.1346(117) & 0.6525(57) & 0.7177(62) \\ \hline
\end{tabular}
\caption{\label{tab:critical_exponents_1} Critical temperatures and main critical exponents for all samples, followed by the critical exponents calculated from experimental values of $\beta$, $\gamma$, and $\delta$ exponents.}
\end{table}

By using the values for the spin dimensionality implied by the HMIM, where the Ising universality class corresponds to n=1 and Heisenberg to n=3, the calculated $\sigma$ are $>3/2$ for all samples and temperatures, corroborating the non-mean field behaviour of the system\cite{fisher_critical_1972}. The $\sigma_H$  constants reach agreement across all samples, at $\sigma_H\approx1.95$. The $\sigma_L$ constants reach agreement across the pellet samples, at $\sigma_L\approx1.91$, although the single crystal constant is close, at $\sigma_L\approx 1.94$. The short-range interaction condition is $\sigma>2-\eta$ (see ref. [\citen{sak_low-temperature_1977}]), with $\eta=\left(0.0363, 0.035\right)$ being the pair correlation exponent for the Ising and Heisenberg universality classes respectively\cite{odor_universality_2004}. We reach an agreement with this limit for the high temperature exponents, and are extremely close in the low temperature exponents, reinforcing our agreement with the short-range interaction universality classes without an interaction constant dependence of the critical exponents. \\ \\

The analysis of the relative contributions of SEI, DMI, and MCA to the changes in magnetic properties with doping required quantification of the FD to HM critical temperature, $T'_C$, as well as the upper and lower critical fields, $H_{C1}$ and $H_{C2}$. These are the critical fields that separate the helical and conical phases, and the conical and field polarised phases, respectively. The relevant critical field points are those at the lowest temperatures available, in our case 4 K. This is because the low temperature value $H_{C1}$ relates directly to the anisotropy constant, K, and that of $H_{C2}$ to the ratio $D^2/J$ (see ref. [\citen{levatic_dramatic_2016}]). A stronger anisotropy will prevent the spins in the helical phase from realigning with the field direction to form the conical phase, while the ratio $D^2/J$ describes the competition between the perpendicular alignment favoured by the DMI and the parallel alignment described by the SEI. Lower values of this ratio therefore result in the field polarised state at lower fields\cite{janson_quantum_2014}. \\ \\

The critical fields are found through the inflection points on the $dM/dH$ curves, as outlined in ref. [\citen{chauhan_origin_2022}]. This is depicted in Fig. S11 of the SI\cite{Supplementary_Information}, where the critical fields for Cu$_2$OSeO$_3$ and Cu$_2$OSe$_{0.9}$Te$_{0.1}$O$_3$ polycrystalline samples are shown, and superimposed onto the low temperature portion of their susceptibility heat-maps. The single crystal sample is omitted, as the existence of singularly oriented easy-axes prevents useful comparison of the anisotropy constant with the polycrystalline samples. \\ \\

The resulting critical fields are shown in table \ref{tab:critical_fields} below. It can be noticed that the $H_{C1}$ values are essentially identical between the samples, while there is significant difference in the $H_{C2}$ values.\\ \\
\begin{table}[ht]
\centering
\begin{tabular}{|l|c|c|c|c|}
\hline
Sample & $H_{C1}$ (Oe) & $H_{C2}$ (Oe)   \\ \hline
Cu$_2$OSeO$_3$    & 315.9        & 843.8           \\ \hline
Cu$_2$OSe$_{0.9}$Te$_{0.1}$O$_3$   & 313.2        & 709.8          \\ \hline
\end{tabular}
\caption{\label{tab:critical_fields} Critical fields for the Cu$_2$OSeO$_3$ and Cu$_2$OSe$_{0.9}$Te$_{0.1}$O$_3$ polycrystalline samples, found through the inflection points of $M-H$ Isotherms.}
\end{table}

This corresponds to a percentage change in the quantity $\sqrt{(H_{C2} T'_C )}$ , and therefore to a change in the DMI constant D, of approximately -9.5\% between the Cu$_2$OSeO$_3$ and 10\%-doped samples. Simultaneously, the change in  $T'_C$ between the two samples of -2.4\% corresponds to the same change in the strength of the exchange parameter, $J$\cite{zivkovic_critical_2014, janson_quantum_2014}. Finally, the change in $H_{C1}$ corresponds to a change in anisotropy constant $K$ of just -0.6\%. The larger change in $D$ with respect to $J$ also agrees qualitatively with previous Lorentz microscopy results in Te-doped single crystals of Cu$_2$OSe$_{0.993}$Te$_{0.007}$O$_3$, where the helical modulation period\cite{kanazawa_noncentrosymmetric_2017}, described by $\lambda \propto J/D$, increased with Te-doping\cite{han_scaling_2020}.

\section*{Conclusion}
We have synthesised and characterised high-purity polycrystalline Cu$_2$OSe$_{1-x}$Te$_x$O$_3$ samples with x = 0, 0.05, and 0.1, and an undoped Cu$_2$SeO$_3$ single crystal. The incorporation of Te into the crystal structure has led to a lattice expansion verified by room-temperature synchrotron pXRD. SANS confirms that Te-doping of the Se site does not break the helimagnetic nature of the system, as shown by the magnetic scattering patterns. We also demonstrate that the conical phase is suppressed and shifted to lower temperatures, along with the skyrmion phase being stable at lower temperatures with the addition of more Te into the crystal structure. We find that the Universality class transition from Heisenberg to Ising like spin dimensionality is robust to this lattice expansion, and there is no significant change to the associated magnetocrystalline anistropy with doping. Instead, Te-doping has the effect of lowering the $H_{C2}$ and $T'_{C}$ critical values, which result from lower DMI and SEI strengths. These findings have implications for future spintronics applications and designing new magnetic materials, while our improved method of HMIM provides a robust technique for understanding the growing number of universality class transitions\cite{chauhan_different_2022, 2023, meng_crossover_2023}. \\ 

\section*{Acknowledgements}

We thank Dr. L. Tan for collecting the room-temperature synchrotron powder X-ray diffraction data at the Australian Synchrotron, a part of the Australian Nuclear Science and Technology Organisation (ANSTO). 
We also would like to thank the ANSTO ACNS sample environment team: G. Davidson, T. d'Adam and C. Baldwin, for their continuous support with the operation of the magnet for our SANS experiment (Proposal No. P15924).
Funding: This research was funded by the Royal Society of New Zealand Marsden Fund (20-UOA-225), AINSE Ltd. Postgraduate Research Award (PGRA) (ALNSTU13239) and the University of Auckland Doctoral Scholarship for M. V\'as. We also acknowledge the support through the Australian Research Council (ARC) through the funding of the Discovery Grant DP170100415 and the funding of the Linkage Infrastructure, Equipment and Facilities Grant LE180100109. This research was undertaken in part on the Powder Diffraction beamline at the Australian Synchrotron, part of ANSTO (Proposal No. PDR21589), as well as on the QUOKKA instrument at the Australian Centre for Neutron Scattering (ACNS) at ANSTO (Proposal No. P15924).

\section*{Author Contributions Statement}

AJF conceived the idea and designed the magnetic experiments. MV synthesised all the samples and performed structural, elemental, and neutron-scattering characterisation. AJF, MFP, SY, and CU performed magnetic measurements. MV, EJV, SY, and EPG performed SANS measurements. AJF and MV analysed the data and wrote the manuscript. All authors contributed to the discussion of the results and the improvement of the manuscript. SY, CU and TS led the project.

\section*{Additional Information}
The authors have no competing interests to declare in this work. 

\section*{Experimental Methods}

Polycrystalline samples of Cu$_2$OSe$_{1-x}$Te$_x$O$_3$ $(0 < x < 0.1)$ were prepared by solid-state sintering. Stoichiometric amounts of high purity CuO (99.5\%), SeO$_2$ (99.99\%) and TeO$_2$ (99.99\%) powders were weighed and mixed into homogenous mixtures before being sealed in evacuated quartz tubes and placed in a tube furnace. The sealed ampules were heated to 610 $\degree$C over 2 hours and held at the sintering temperature for 48 hours before being slowly cooled to room temperature over 6 hours. 
The samples were pressed into 10 mm diameter pellets, sealed in an evacuated quartz ampule and resintered at 550 $\degree$C for 24 hours. In order to fit into the SQUID’s bore, these were then reshaped into cylindrical ellipses. \\

Single crystals were grown by chemical vapour transport using NH$_4$Cl as the transport agent\cite{panella_seeded_2017}. Stoichiometric amounts of high purity CuO (99.5\%) and SeO$_2$ (99.99\%) powders were weighed and mixed into a homogeneous mixture along with 0.4 mg/cm$^3$ of NH$_4$Cl before being sealed in an evacuated quartz tube. The sealed ampule was positioned in a two-zone furnace with the source powder located in the hotter source zone and the empty side in the cooler sink zone. The source and sink zones were heated at 5 $\degree$C/min to 640 $\degree$C and 540 $\degree$C, respectively. This temperature gradient was maintained for five weeks. The source zone was cooled to room temperature seven hours before the sink zone to allow for the inversion of the temperature gradient to allow for the transport agent to deposit on the source powders instead of on the crystals. The crystals were cleaned with ethanol in an ultrasonic bath to remove any remaining NH$_4$Cl residue on their surfaces. The yielded dark green single crystals ranged in size between 1-2 mm$^3$. \\ \\

The structural properties of the polycrystalline samples were determined using Rietveld refinements performed using the FullProf software\cite{RietveldH.M.1969Aprm,RODRIGUEZCARVAJAL199355} on the room-temperature synchrotron X-ray powder diffraction data collected on the Powder Diffraction Beamline at the Australian Synchrotron .Data were collected using 21 keV X-rays with a wavelength of 0.59053 {\AA} in two scans of 150 seconds spliced together to account for gaps along 2$\theta$ in the detector. Additional chemical analysis was done using electron dispersive spectroscopy (EDS), which was carried out on an FEI Quanta 200F Environmental Scanning Electron Microscope (ESEM) using an EDAX Genesis Energy Dispersive X-ray Analysis system, counting for 100 live seconds. All SEM images were taken in backscattered electron mode using a solid-state detector to collect EDS spectra for elemental analysis of the polycrystalline samples. All images used a 20 kV electron beam with a spot size of 4 mm at a working distance of approximately 10 mm in ESEM mode. Prior to imaging, all three samples were coated with an approximately 20 nm thick layer of Pt metal to improve conductivity and imaging. This was done using a Q150RS Sputter Coater via applying a 20 mA current for 60 seconds. \\\\

The SANS experiments were conducted on QUOOKA, the monochromatic small-angle neutron-scattering instrument at Australian Centre for Neutron Scattering (ACNS), ANSTO \cite{GILBERT20061180,Wood:uh5003} (Sydney, Australia). For all measurements, approximately 0.5 g of each polycrystalline Cu$_2$OSe$_{1-x}$Te$_x$O$_3$ $(0\leq x \leq 0.1)$ sample was filled into an opened circular quartz Hellma cells with a diameter of 1.5 cm and pathlength of 1 mm. Immediately prior to the study, each sample was placed in an aluminium holder containing a bespoke cavity into which the cuvette could be located on the end of the cryomagnet stick. Measurements were performed at equal source-to-sample and sample-to-detector distances of 20 m with 5 {\AA} and 10\% wavelength resolution, a source aperture diameter of 50 mm and a 7.5 mm Cd sample aperture. An Oxford Instruments Spectromag horizontal superconducting magnet with a peak field of 10 T was used to apply a magnetic field perpendicular to the beam direction for the experiment to observe all the magnetic phases: Helical, Conical, and Skyrmion. For all the measurements, a sample stabilisation time of 300 seconds was incorporated to ensure temperature accuracy. The same protocol was used for all measurements: The sample was zero field cooled (ZFC) to base temperature (4 K) before any magnetic fields were applied and the temperature sweeps were carried out. To reset the magnetic system, the magnetic field was driven to zero (if one was applied), before heating the sample to 120 K before ZFC back to base for the next measurement. Temperature sweep measurements had the applied magnetic field set either at 0 T or 0.02 T before heating up in incremental temperature steps from 4 K to 60 K. All the individual scans were measured for 5 minutes except for the 0 T temperature sweep for Cu$_2$OSeO$_3$ which was done as 10 individual 60-second scans. A 5-minute time period was chosen to readily observe clear peak intensity above background. \\

Data were reduced using the NIST NCNR SANS reduction macros modified for QUOKKA \cite{Kline:Igor}, using the Igor Pro 7 software package (Wavemetrics, Lake Oswego, OR) with data corrected for detector sensitivity and using a 120 K scan as the background. Data were reduced using two approaches: (i) radially as a function of the magnitude of the scattering vector, Q, and (ii) as a function of angle with respect to the applied field direction within a defined Q range (i.e. annular reduction).
Due to packing density variations between polycrystalline samples, slight variations in intensity around the beamstop were observed; to account for this, and to avoid re-positioning the beamstop for each measurement, a mask was applied during the data reduction to remove the high intensity of unscattered neutrons observed in the middle of the detector images. Data was analysed by fitting the resulting Gaussian peaks, yielding the maximum peak intensity in the annular data and the Q-position of the peaks in the radial data, with a flat baseline. \\ \\

DC Magnetometry was performed on a SQUID magnetometer (Quantum Design MPMS3) with the VSM option. The single crystal and pellet samples were attached to quartz paddles using Marubu fixogum. Initial magnetisation-temperature measurements were performed by cooling to 4K from RT in zero field before heating and measuring from 40-70 K at 0.3K/min, still at zero field. Slight remnant fields in the superconducting magnet can affect the magnetisation. The raw data are fitted with a cubic spline, which is then differentiated to find the critical temperatures. \\ \\
The magnetisation-field sweeps used for the modified Arrott plots and scaling analysis are then centred on the FD to PM transition to the closest 0.5 K. Each sweep consisted of zero-field cooling from 300 K to 4 K, zero-field warming to the temperature of the isotherm, before taking a sweep from 0 to 7 T. The initial 2100 Oe was measured in 10 Oe increments, followed by 1000 Oe increments up to 70 kOe . Finally, the sample was heated to 300K to remove the magnetic ordering and the external magnet was demagnetised in an oscillating fashion from 15 kOe, to reduce the remnant magnetic field in the next cooling phase of the $M-H$ runs. These magnetisation-field sweeps were corrected for the demagnetising field of the sample. \\ \\ 
Additional magnetisation-field sweeps, used for determining critical field at low temperature and for producing the susceptibility heat-maps, were performed on 0\% and 10\% Te-doped powder capsules. The zero-field cooling and warming steps are the same as above, with measurement temperatures from 4 K to 64 K in 1 K increments. At each temperature, the field sweep was performed from 0 Oe to 1500 Oe in 0 Oe intervals. As the form of the demagnetising factor is not known for these samples, these sweeps are not corrected. This has a significantly lower impact at the lower fields used for this section of the analysis, as the magnitude of the magnetisation is lower. \\ \\

\newpage

\bibliography{Paper_Skyrmion_Universality_Classes}

\clearpage

\newpage

\newpage

\newpage

\newpage

\newpage

\newpage

\newpage

\newpage

\newpage

\section*{Supplementary Information: Universality  Class Transition Across the Helimagnetic Ordering in Te-doped Cu$_2$OSeO$_3$}

\setcounter{figure}{0}
\newpage

\section*{Synchrotron pXRD}
\begin{figure}[ht]
\centering
\includegraphics[width=\linewidth]{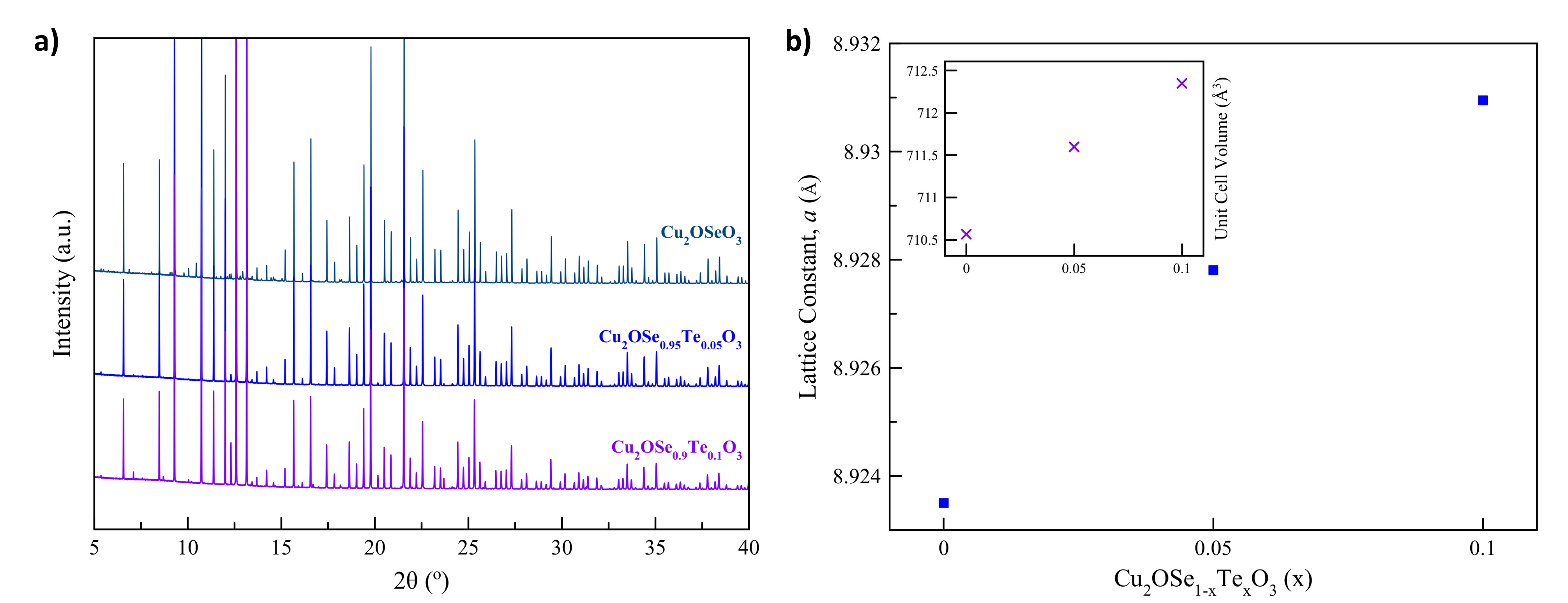}
\caption{a) Room temperature synchrotron pXRD patterns of polycrystalline Cu$_2$OSe$_{1-x}$Te$_x$O$_3$ samples where x = 0, 0.05 and 0.1. b) Lattice constants and unit cell volumes for the corresponding samples.} 
\label{fig:Synchrotron pXRD}
\end{figure}
\newpage

\section*{Rietveld Refinements}
\begin{figure}[h!]
\centering
\includegraphics[height=1.2\linewidth]{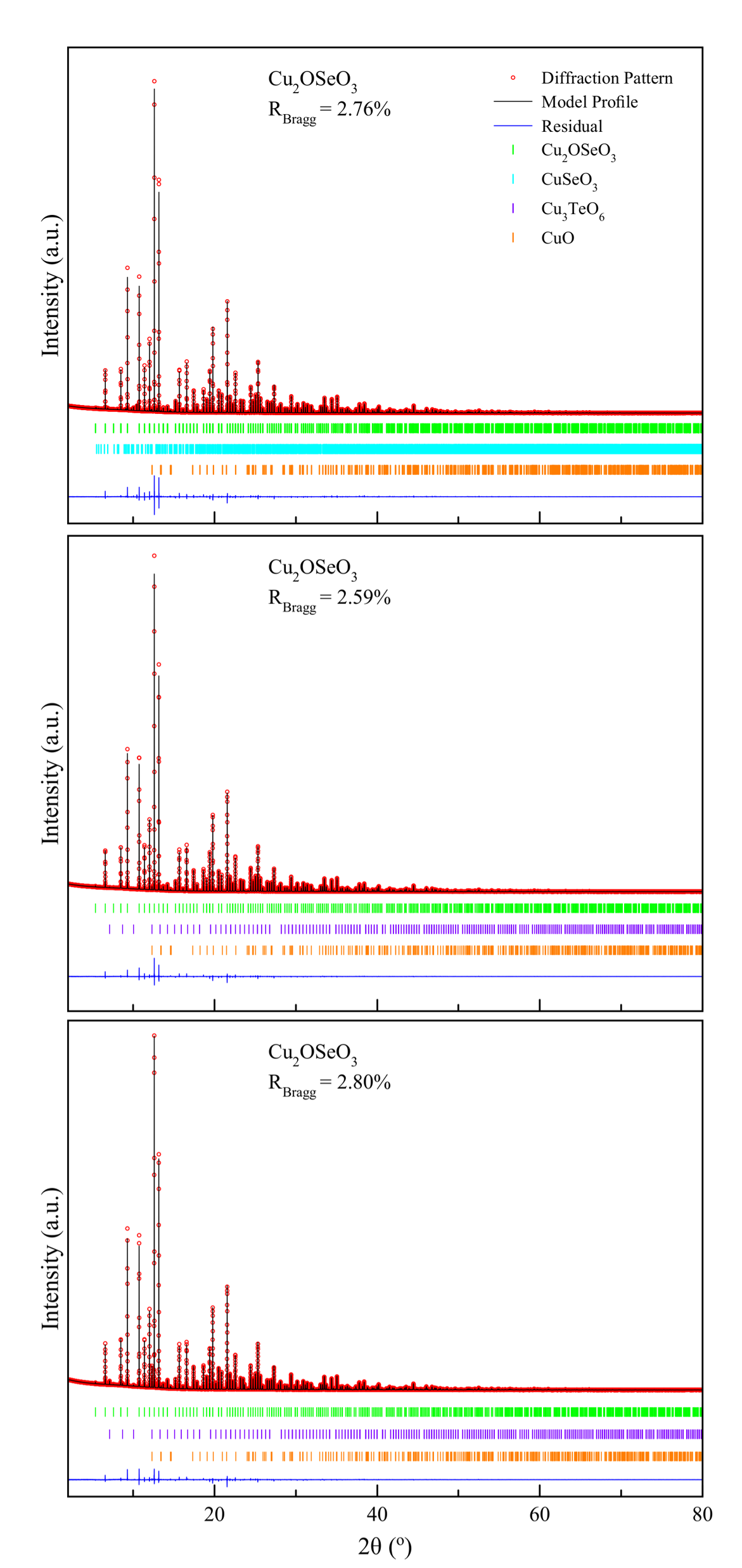}
\caption{Rietveld refinements of the refined room temperature synchrotron pXRD patterns of polycrystalline Cu$_2$OSe$_{1-x}$Te$_x$O$_3$ samples where x = 0, 0.05, and 0.1.} 
\label{fig:Rietveld Refinements}
\end{figure}
\newpage

\section*{Elemental Analysis}

\begin{figure}[h]
\centering
\begin{subfigure}{.45\textwidth}
  \centering
  \includegraphics[width=1\linewidth]{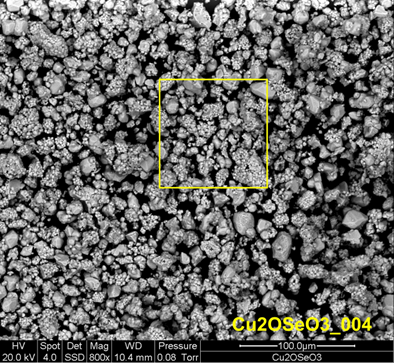}
  
\end{subfigure} 
\begin{subfigure}{.45\textwidth}
  \centering
  \includegraphics[width=1\linewidth]{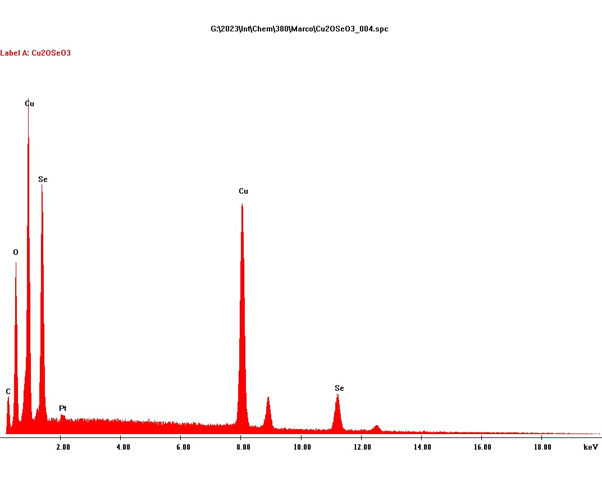}
  
\end{subfigure}
\caption{A backscattered SEM image of undoped polycrystalline Cu$_2$OSeO$_3$ along with the EDS spectrum.}
\label{fig:EDS_Undoped}
\end{figure}

\begin{figure}[h]
\centering
\begin{subfigure}{.45\textwidth}
  \centering
  \includegraphics[width=1\linewidth]{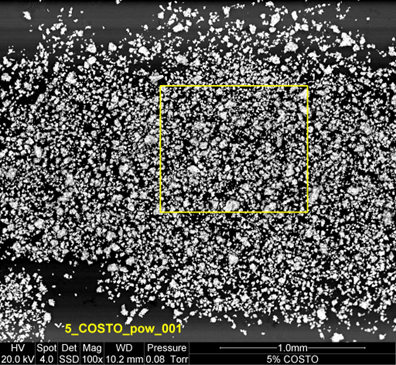}
  
\end{subfigure} 
\begin{subfigure}{.45\textwidth}
  \centering
  \includegraphics[width=1\linewidth]{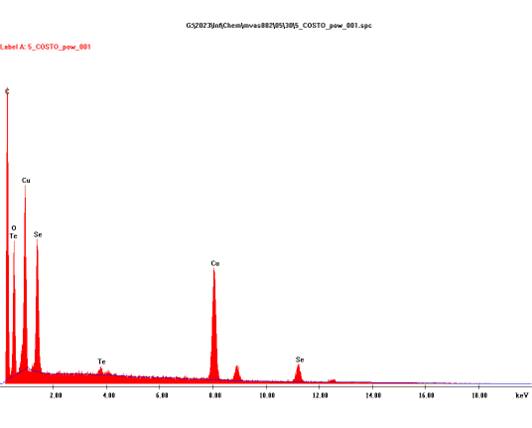}
  
\end{subfigure}
\caption{A backscattered SEM image of Te-doped polycrystalline Cu$_2$OSe$_{0.95}$Te$_{0.05}$O$_3$ along with the EDS spectrum.}
\label{fig:EDS_5}
\end{figure}

\begin{figure}[h]
\centering
\begin{subfigure}{.45\textwidth}
  \centering
  \includegraphics[width=1\linewidth]{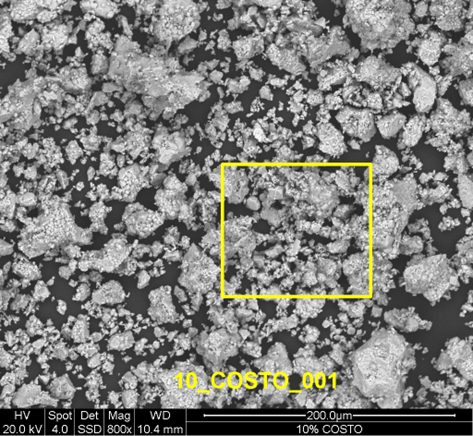}
  
\end{subfigure} 
\begin{subfigure}{.45\textwidth}
  \centering
  \includegraphics[width=1\linewidth]{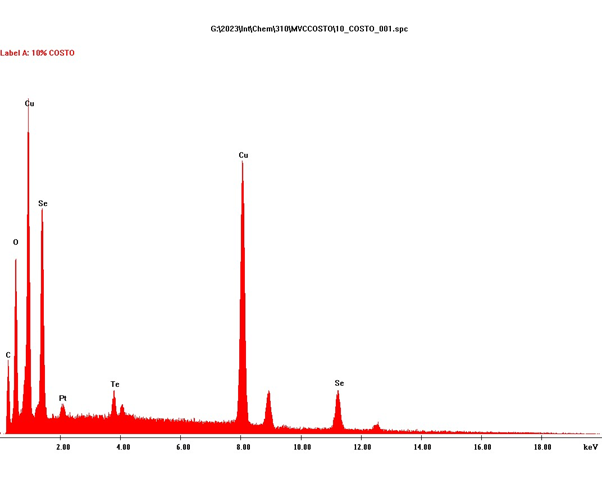}
  
\end{subfigure}
\caption{A backscattered SEM image of Te-doped polycrystalline sample Cu$_2$OSe$_{0.9}$Te$_{0.1}$O$_3$ along with the EDS spectrum.}
\label{fig:EDS_10}
\end{figure}

\begin{table}[h]
\centering
\begin{tabular}{|c|cccc|c|}
\hline
\multirow{2}{*}{Composition (x)} & \multicolumn{4}{c|}{Atomic \%}                                                              & \multirow{2}{*}{Te-Doping (\%)} \\ \cline{2-5}
                                 & \multicolumn{1}{l|}{O K}   & \multicolumn{1}{l|}{Cu K}  & \multicolumn{1}{l|}{Se K}  & Te L &                                 \\ \hline
0                                & \multicolumn{1}{l|}{43.40} & \multicolumn{1}{l|}{37.11} & \multicolumn{1}{l|}{19.48} & 0.00 & 0.00                            \\ \hline
0.05                             & \multicolumn{1}{l|}{51.97} & \multicolumn{1}{l|}{31.88} & \multicolumn{1}{l|}{15.46} & 0.69 & 4.46                            \\ \hline
0.10                             & \multicolumn{1}{l|}{40.27} & \multicolumn{1}{l|}{40.27} & \multicolumn{1}{l|}{20.98} & 1.33 & 6.34                            \\ \hline
\end{tabular}
\caption{\label{tab:EDS_Data}EDS data collected for polycrystalline Cu$_2$OSe$_{1-x}$Te$_x$O$_3$ samples quantified through the O, Cu, and Se K-edges, and the Te L-edge. The determined Te occupation of the Se sites is shown in terms of percentage.}
\end{table}

\clearpage

\section*{Extended SANS Data}

\begin{figure}[h!]
\centering
\includegraphics[width=0.9\textwidth]{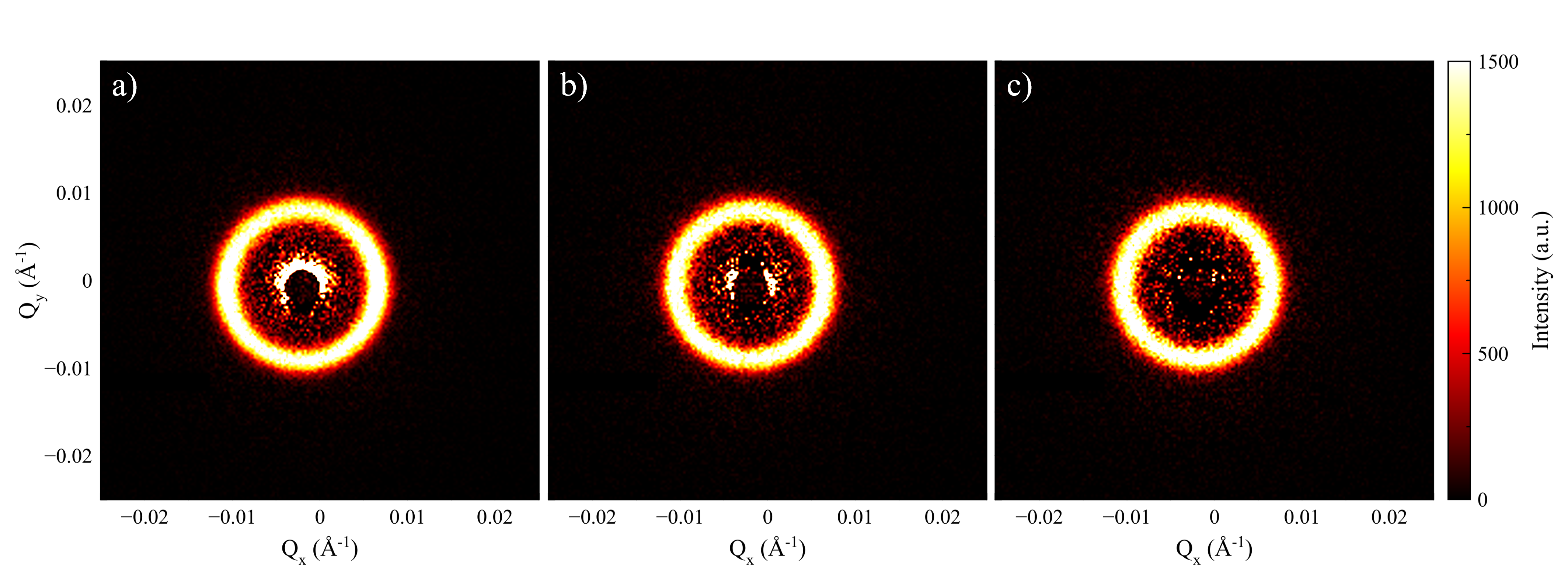}
\caption{Polycrystalline SANS detector images at 4 K  and 0 Oe displaying the helical pattern, in the samples a) Cu$_2$OSeO$_3$, b) Cu$_2$OSe$_{0.95}$Te$_{0.05}$O$_3$, c) Cu$_2$OSe$_{0.9}$Te$_{0.1}$O$_3$.} 
\label{fig:SI SANS Helical}
\end{figure}

\begin{figure}[h!]
\centering
\includegraphics[width=0.9\textwidth]{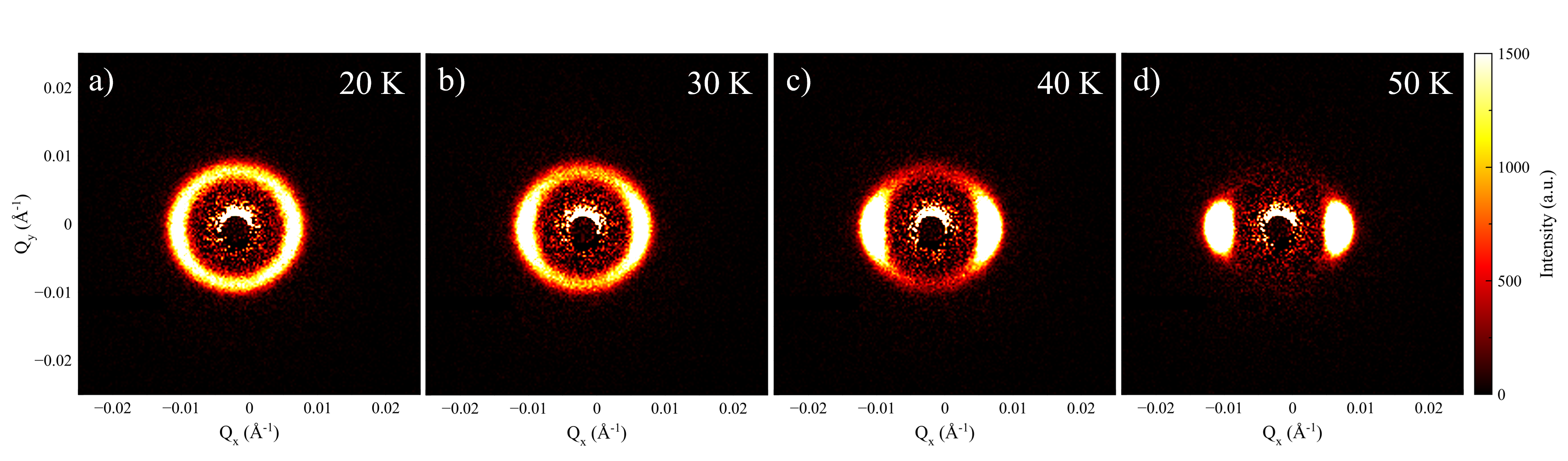}
\caption{Polycrystalline SANS detector images across the helimagnetic to conical transition in undoped Cu$_2$OSeO$_3$. Patterns were taken at 200 Oe, at temperatures: a) 20 K, b) 30 K, c) 40 K, d) 50 K.} 
\label{fig:SI SANS H2C}
\end{figure}

\begin{figure}[h!]
\centering
\includegraphics[width=0.9\textwidth]{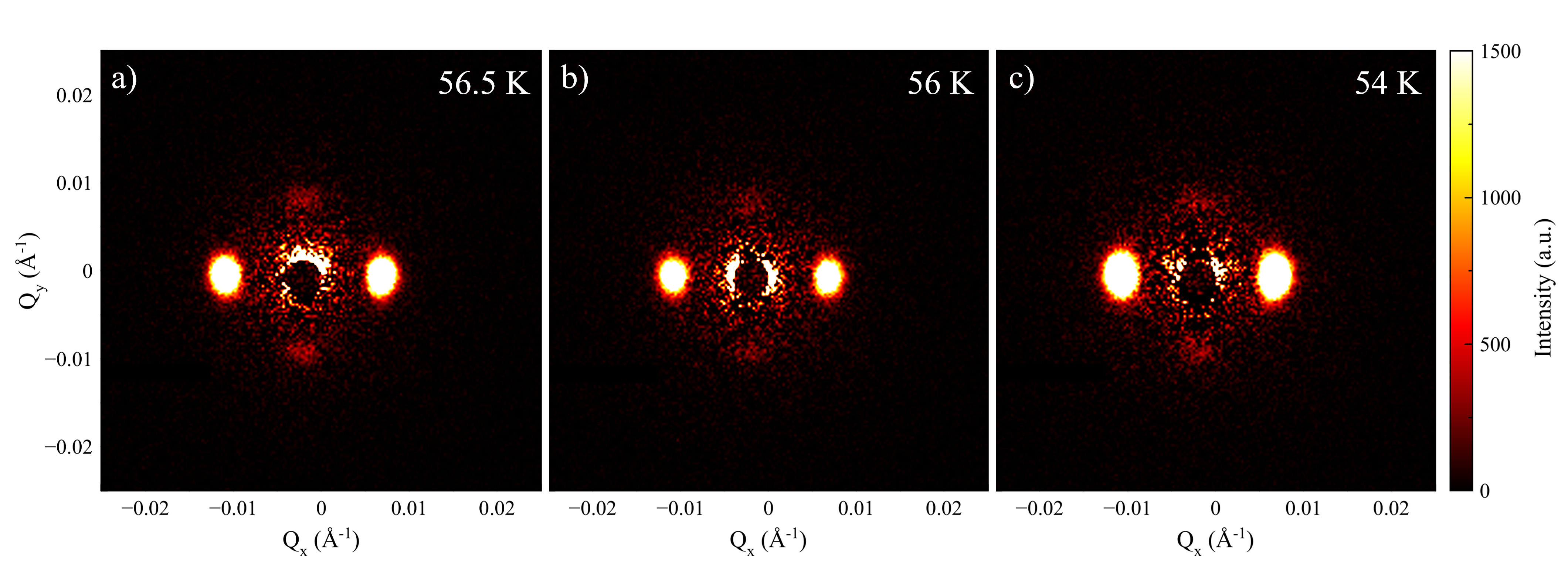}
\caption{Polycrystalline SANS detector images at 200 Oe, displaying the skyrmion and conical patterns at the temperatures: a) 56.5 K for Cu$_2$OSeO$_3$, b) 56 K for Cu$_2$OSe$_{0.95}$Te$_{0.05}$O$_3$, c) 54 K for Cu$_2$OSe$_{0.9}$Te$_{0.1}$O$_3$.} 
\label{fig:SI SANS skyrmion}
\end{figure}

\begin{figure}[h!]
\centering
\includegraphics[width=0.9\textwidth]{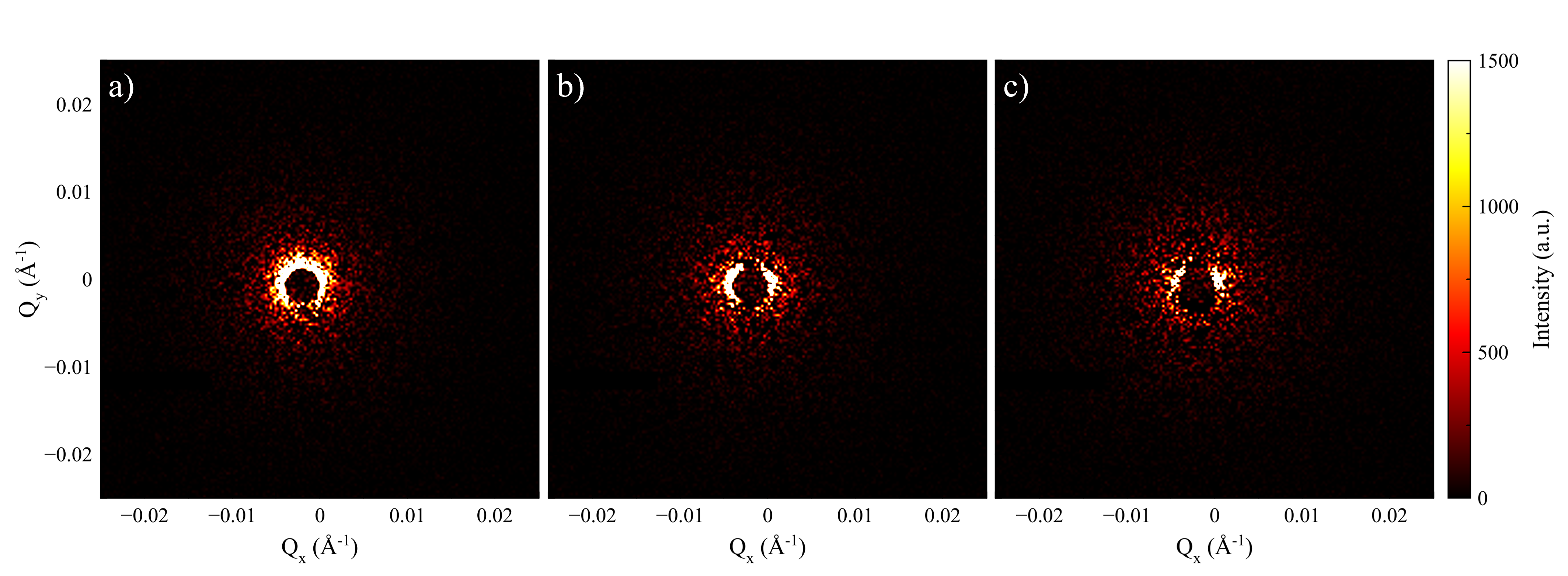}
\caption{Polycrystalline SANS detector images in the fluctuation disordered regime. Patterns were taken at 0 Oe, at the temperatures: a) 60 K for Cu$_2$OSeO$_3$, b) 60 K for Cu$_2$OSe$_{0.95}$Te$_{0.05}$O$_3$, c) 60 K for Cu$_2$OSe$_{0.9}$Te$_{0.1}$O$_3$. This confirms the lack of fluctuation disorder ring, which agrees with the Wilson-Fisher scenario.} 
\label{fig:SI SANS FD}
\end{figure}
\clearpage

\section*{Demagnetising Field Correction}
The demagnetising field is corrected in all magnetic measurements shown in this work. This is the field produced by a magnetised sample, which is arranged opposite to the external field, therefore reducing the internal field experienced by the sample itself. It is this internal field which drives the phase transitions seen across the field sweeps, and as such it is important to perform this correction. We approximate the samples as general ellipsoids, where the field is aligned along the a semi major axis with a>b>c. The demagnetising factor N, such that $H_{int}=H_{ext}-NM$, can then be calculated using the method outlined in ref. [\citen{osborn_demagnetizing_1945}] for a general ellipsoid with field aligned to axis a. The pellet samples all have approximately the same ratio of axis lengths, resulting in N$\approx0.139$. The single crystal is approximately spherical, resulting in N$\approx1/3$.  

\newpage

\section*{Justification of Heat-Map MIM}
Rather than following the Modified Iteration Method (MIM) outlined in Chauhan \textit{et al.}\cite{chauhan_different_2022}, we have computed the sum of squared residuals of the normalised slopes across the parameter space. The modification of the MIM is due to the nature of the Arrott-Noakes relation\cite{arrott_approximate_1967}, $[H/M]^{1/\gamma}=c+SM^{1/\beta}$, which is used to fit the high field sections of the Modified Arrott Plots (MAPs) with S and c as free parameters. The normalised slope of a given temperature $T_1$ was then computed as $S(T_1 )/S(T_C ) =M_1(H)^{[\gamma+\beta]/\gamma\beta}-c_1 M_1(H)^{1/\gamma}/\left[M_C (H)^{[\gamma+\beta]/\gamma\beta}-c_C M_C (H)^{1/\gamma}\right]$ where the magnetisations and y intercepts of temperature $T_1$ and $T_C$ are denoted by their respective subscripts. As the y intercept at the critical temperature, $c_C$, is zero, and $M_1(H)^{[\gamma+\beta]/\gamma\beta}/M_C(H)^{[\gamma+\beta]/\gamma\beta}\gg  c_1 M_1(H)^{1/\gamma}/M_C(H)^{[\gamma+\beta]/\gamma\beta}$  for small $c_1$, the normalised slopes are not dependent on the individual $(\beta,\gamma)$ but the ratio $[\gamma+\beta]/\gamma\beta$. As a result, rather than a single set of exponents that result in a minimum in the squared residual of normalised slopes, there exists a set of minima along an isoline corresponding to a constant $[\gamma+\beta]/\gamma\beta$ value, and we require the use of the $\delta_C$ value to select a unique solution amongst the set of equivalent pairs through the Widom scaling relation (eq. 5 in the main text). \\ \\

\begin{figure}[ht]
\centering
\includegraphics[width=\linewidth]{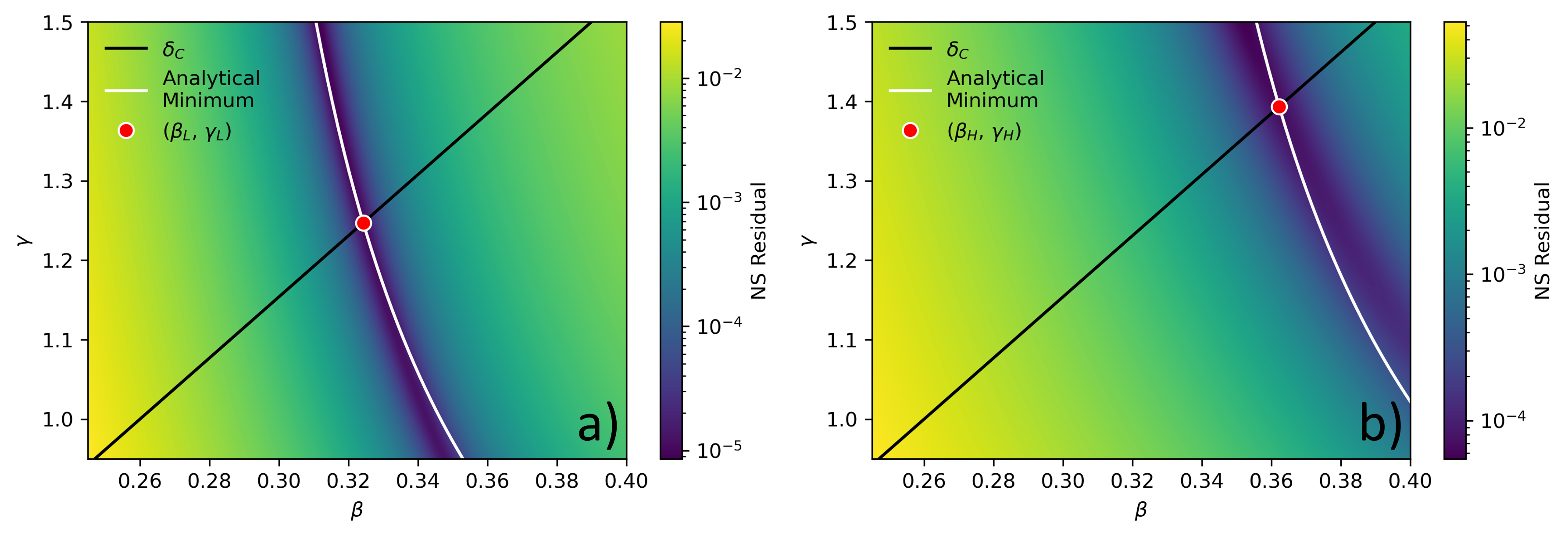}
\caption{Normalised slope heat-maps for Cu$_2$OSeO$_3$ Pellet sample. a) is below $T_C$, b) is above $T_C$}
\label{fig:Heat_Map_Analytical_Min}
\end{figure}

This is best demonstrated in Fig. \ref{fig:Heat_Map_Analytical_Min}, where the calculated isolines of both the above and below $T_C$ datasets are shown as white lines. These isolines diverge from the experimental minima both at low values of $\gamma$ and high values of $\beta$, as the $c_1 M_1 (H)^{1/\gamma}/M_C (H)^{[\gamma+\beta]/\gamma\beta}$ term becomes more significant. This is accompanied by a widened region of minima due to the temperature dependence of the $c_1$ term. However, there remains a set of minima from which our exponents must be selected, necessitating the use of the $\delta_C$ exponent as a constraint. 
\newpage

\section*{Critical Fields}

\begin{figure}[ht]
\centering
\includegraphics[width=\linewidth]{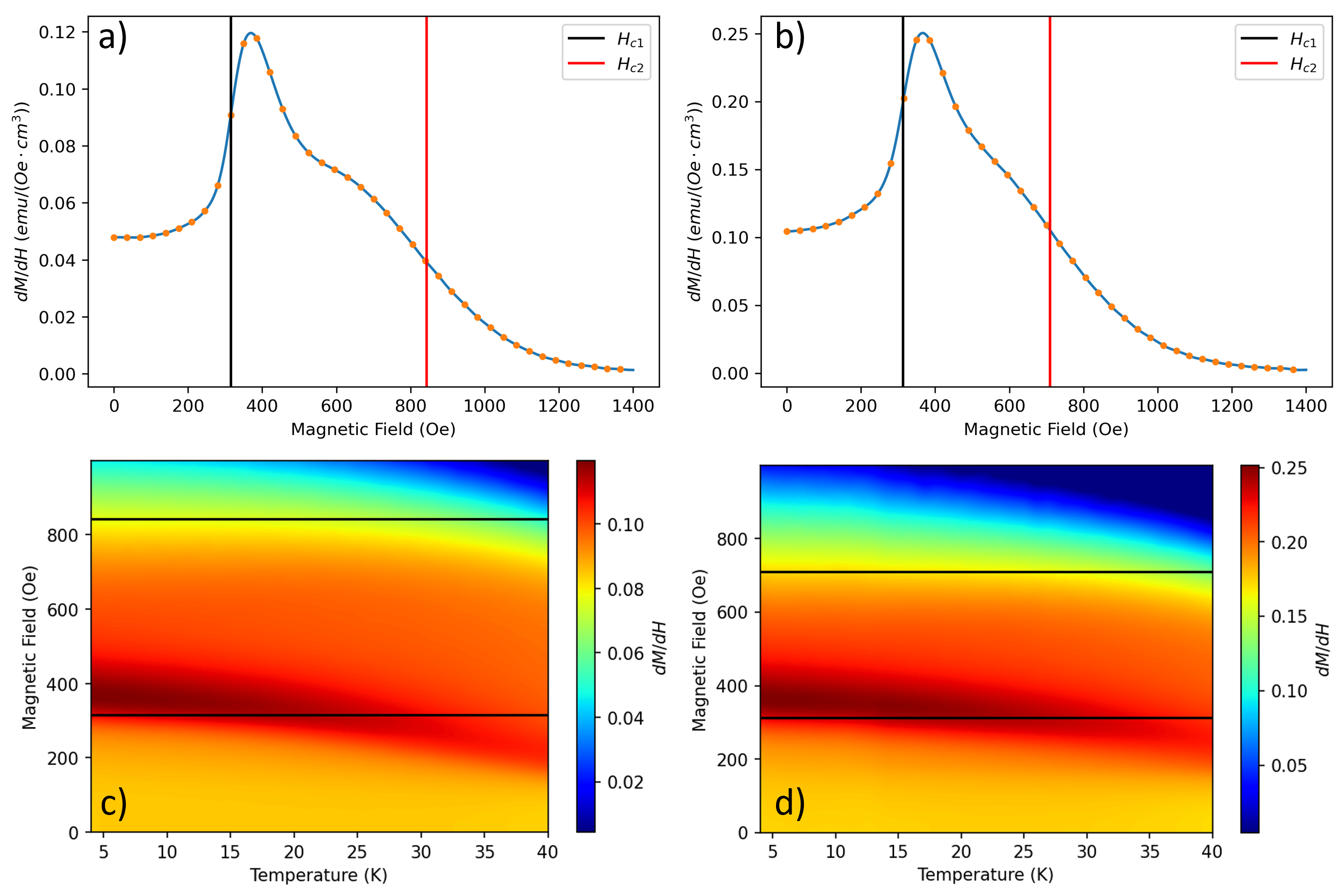}
\caption{Critical fields of the polycrystalline samples. (a),(b): Derivatives of the $M-H$ Isotherms at 4 K, showing the $H_{C1}$ and $H_{C2}$ values at the inflection points of the 0\% and 10\% samples respectively. $H_{C1}$ is shown in black and $H_{C2}$ in Red, while the blue line is a spline fit to the derivative. (c),(d): $H_{C1}$ and $H_{C2}$ superimposed onto the phase diagrams of the samples corresponding to the figures above.}
\label{fig:Critical_Fields}
\end{figure}

Shown in Fig. \ref{fig:Critical_Fields} are the gradients of the magnetisation with respect to magnetic field, with the associated critical fields for the polycrystalline samples. These are quantified in the main text in table 3.

\end{document}